\documentclass[twocolumn,prb,aps,showpacs,floatfix,superscriptaddress,preprintnumbers,amsmath,amssymb]{revtex4}

\usepackage{graphicx}

\begin{document}

\renewcommand{\floatpagefraction}{0.5}

\title{Time Resolved Stroboscopic Neutron Scattering of Vortex Lattice Dynamics in Superconducting Niobium}

\author{S. M\"uhlbauer}

\affiliation{Technische Universit\"at M\"unchen, Physik Department E21, D-85748 Garching, Germany}

\affiliation{Neutron Scattering and Magnetism Group, Institut for Solid State Physics, ETH Z\"urich, Z\"urich, Switzerland}

\affiliation{Forschungsneutronenquelle Heinz Maier-Leibnitz (FRM II), D-85748 Garching, Germany}

\author{C. Pfleiderer}

\affiliation{Technische Universit\"at M\"unchen, Physik Department E21, D-85748 Garching, Germany}

\author{P. B\"oni}

\affiliation{Technische Universit\"at M\"unchen, Physik Department E21, D-85748 Garching, Germany}

\author{ E. M. Forgan}

\affiliation{School of Physics and Astronomy, University of Birmingham, Birmingham, UK}

\author{ E. H. Brandt}

\affiliation{Max Planck Institut f\"ur Metallforschung, Stuttgart, Germany}

\author{A. Wiedenmann}

\affiliation{Institut Laue Langevin ILL, Grenoble, France}

\author{U. Keiderling}

\affiliation{Helmholtz Zentrum Berlin, BENSC, D-14109 Berlin, Germany}

\author{G. Behr}

\affiliation{Leibnitz-Institut f\"ur Festk\"orper- und Werkstoffforschung IFW, D-01069 Dresden, Germany}

\date{\today}

\begin{abstract}
Superconducting vortex lattices, glasses and liquids attract great interest as model systems of crystallization and as a source of microscopic information of the nature of superconductivity. We report for the first time direct microscopic measurements of the vortex lattice tilt modulus $c_{44}$ in ultra-pure niobium using time-resolved small angle neutron scattering. Besides a general trend to faster vortex lattice dynamics for increasing temperatures we observe a dramatic changeover of the relaxation process associated with the non-trivial vortex lattice morphology in the intermediate mixed state. This changeover is attributed to a Landau-branching of the Shubnikov domains at the surface of the sample. Our study represents a showcase for how to access directly vortex lattice melting and the formation of vortex matter states for other systems.
\end{abstract}


\vskip2pc

\maketitle

\section{Introduction and Motivation}
The morphology of superconducting vortex lattices (VL) attracts great interest as a source of microscopic information of the nature of the superconductivity and as model systems of condensed matter. The elastic matrix $\Phi_{\alpha \beta}$ of a VL thereby describes the energy associated with a distortion of the VL due to thermal fluctuations, gradients of magnetic field or temperature, pinning and the presence of transport currents. Analogous to crystal lattices the elastic matrix $\Phi_{\alpha \beta}$ of a VL determines the thermal stability and the state of aggregation of superconducting vortex matter: Besides the regular Abrikosov VL, VL Bragg glasses, liquids and ices have been identified \cite{Klein:01, Maple:03, Li:04}.

At VL melting transitions the shear modulus discontinuously jumps to $c_{66}=0$, where the long range order vanishes. VL melting was observed for various superconducting systems, mostly for high-$T_c$ compounds due to their high transition temperature \cite{Cubitt:93, Vinokur:90}, but also in compounds characterized by disorder as NbSe$_2$ and MgB$_2$ \cite{Klein:01, Maple:03, Li:04}. Surprisingly, a melting transition was recently reported also for the heavy fermion compound URu$_2$Si$_2$ \cite{Okazaki:08} in the clean limit. Melting transitions show up as characteristic dips of the differential resistivity and can also lead to tiny jumps of the local magnetization of the order of few tenths of a Gauss, detectable with sensitive Hall probes \cite{Zeldov:95}. Measurements to detect melting transitions are intricate, as effects induced by depinning can yield similar results. However, the presence of pinning is required in general, as a perfect, pinning free VL shows no signature in the resistivity at the melting transition \cite{Brandt:95}.

Moreover the elastic matrix $\Phi_{\alpha \beta}$ of VLs is intimately related to the pinning and depinning properties of superconducting vortices. This is especially important for technical applications: If transport currents are applied to superconducting materials, the Lorentz force acting on the vortices leads --- with increasing current --- to dissipative processes such as vortex creep \cite{Feigel:89}, thermally assisted flux-flow (TAFF) \cite{Kes:89} and flux-flow (FF) \cite{Kim:65}. Therefore, the ability of superconducting materials to carry large transport currents for technical applications is closely connected to the pinning properties of superconductors and the elasticity of the VL. The elastic constants of the VL $c_{11}$ for compression, $c_{44}$ for tilt and $c_{66}$ for shear hence reflect the microscopic nature of the superconductivity as well as impurity or surface properties of the superconducting sample due to pinning \cite{Andrej:07, Li:06, Andrej:03, Xiao:99}.

The experimental access to the elastic matrix $\Phi_{\alpha \beta}$ of VLs --- in particular for non-equilibrium states --- by macroscopic bulk techniques such as the transport properties \cite{Andrej:07}, the magnetization \cite{Zeldov:95} or measurements using vibrating reeds \cite{Xu:92} is strongly influenced by parasitic pinning effects as well as geometric effects. Moreover, the momentum-dependence of the elasticity of VLs cannot be determined unambiguously by macroscopic measurements. Due to the use of thin film samples, microscopic surface sensitive techniques such like decoration or magneto-optical methods suffer from similar or even stronger pinning and geometry induced effects. In contrast, local probes such as muon spin relaxation ($\mu$SR) and scattering techniques such as neutron scattering yield microscopic information on bulk VLs. However, as the accessible timescale of inelastic neutron scattering techniques is still too short for VL dynamical properties, neutron scattering was up to now limited to characterize the static properties of VLs with only a few exceptions \cite{Charalambous:02, Thorel:73, Forgan:2000, Pautrat:2003}.

In this paper we report direct microscopic measurements of the VL tilt modulus $c_{44}$ with drastically reduced limitations due to surface pinning in ultra-pure bulk Niobium (Nb) using a time-resolved neutron scattering technique as combined with a tailored magnetic field setup. With its low Ginzburg-Landau parameter $\kappa$, situated close to the border of type-I and type-II behaviour, the superconductivity in Nb is ideally suited as model system for systematic studies of vortex matter \cite{Laver:09,Laver:09b, Laver:06, Muehlbauer:09}. By imposing a periodic tilting of the magnetic field, we induce a relaxation process of the VL which can be described by a diffusion process in the limit of uniform tilt. The diffusion constant of this diffusion process is given by the tilt modulus $c_{44}$ of the VL and the flux flow resistivity $\rho_{\rm FF}$. The characteristic properties of the diffusion process are observed by means of time resolved stroboscopic small angle neutron scattering (SANS) \cite{Wiedenmann:06}. The relaxation processes observed show increasing VL stiffness with increasing magnetic field $H$ and reduced damping with increasing temperature $T$. This behaviour agrees well with calculations performed within a VL diffusion model \cite{Brandt:90}. Besides these general trends, we observe a dramatic change of the relaxation processes associated with the non-trivial VL morphology in the intermediate mixed state (IMS). 

Our study represents a showcase for how to access directly VL melting, the formation of vortex glass states and slow vortex dynamics also in unconventional superconductors, notably the cuprates, heavy-fermion systems, borocarbide or ironarsenide systems.

The outline of this paper is as follows: In the following section (\ref{VL_elasticity}), we briefly revisit the salient features of the elasticity of superconducting VLs. The experimental setup which was developed for our measurements is discussed in depth in section (\ref{Experimental_setup}). The results obtained for the tilt modulus $c_{44}$ are presented in section (\ref{Results_stroboscopic}). The response of the VL to a changed magnetic field environment in the $k=0$ limit with a diffusion Ansatz \cite{Brandt:90, Brandt:95, Brandt:91, Kes:89} is discussed in section (\ref{interpretation}). Finally, the relevance of our experimental setup for the investigation of different magnetic systems is discussed in section (\ref{conclusion}).

\section{Elasticity of Vortex Lattices}
\label{VL_elasticity}
In the following, we introduce the elastic properties of VLs. We first concentrate on the derivation of the free energy of an arbitrarily curved VL using a Ginzburg-Landau ansatz. We then derive the elastic matrix $\Phi_{\alpha \beta}$ and deduce the eigenfrequencies, characteristic timescales and the moduli for compression, tilt and shear of superconducting VLs. In particular, we discuss the elastic moduli for uniform tilt which are relevant for the interpretation of our measurements of the VL elasticity in superconducting Nb. For a detailed theoretical description on VL elasticity, we refer to \cite{Brandt:90, Brandt:95, Brandt:91, Kes:89}.

\subsection{Elastic Energy of Vortex Lattices}
\begin{figure}
\begin{center}
\includegraphics[width=0.5\textwidth]{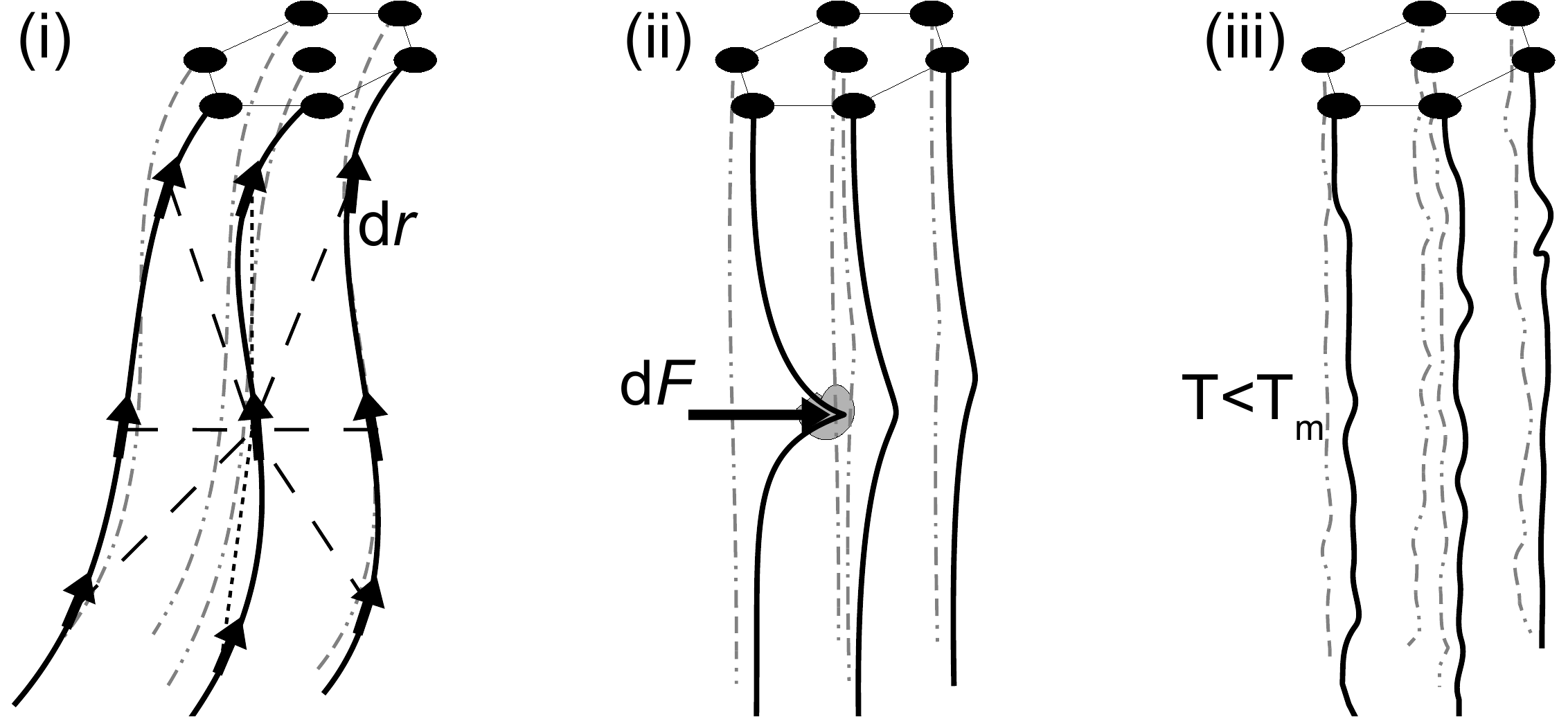}
\caption{Panel (i) schematically depicts the inter (dotted) and intra (dashed) vortex interactions determining a particular symmetry, structure and elasticity of VLs. Panel (ii) depicts the cusp-like response of a superconducting VL to a pinning force $dF$. Panel (iii) schematically depicts a thermally fluctuating superconducting VL for a temperature below the melting temperature $T_m$.}
\label{Vl_elast_struc_2}
\end{center}
\end{figure}

Similar to the elasticity of crystal lattices, which is determined by electrostatic or covalent forces, the elasticity of superconducting VLs is determined by the vortex-vortex interactions.
The free energy $F$ of an arrangement of arbitrarily curved vortices in the Ginzburg-Landau regime close to $T_c$ has been approximated by Brandt \cite{Brandt:95} and can be written in terms of two contributions
\begin{equation}
\begin{split}
& F(r_i\{z\}) = \frac{\phi_0^2}{8\pi \lambda^2 \mu_0}\sum_i \sum_j \left( \int d{\bf r}_i \int d{\bf r}_j \frac{e^{-r_{ij}/\lambda'}}{r_{ij}}- \right.\\
& \left.\int \left|d{\bf r}_i\right| \int \left| d{\bf r}_j \right| \frac{e^{-r_{ij}/\xi'}}{r_{ij}} \right)\,\,
\end{split}
\label{arb_curved_vl}
\end{equation}
with
\begin{equation}
r_{ij}=|{\bf r}_i -{\bf r}_j|,
\end{equation}
\begin{equation}
\lambda'=\lambda / \langle |\psi|^2 \rangle ^{1/2} \sim \lambda/(1-b)^{1/2},
\end{equation}
\begin{equation}
\xi'=\xi/[2(1-b)]^{1/2}.
\end{equation}
and the reduced field $b=B/B_{c2}$. The first term of the sum in eq. (\ref{arb_curved_vl}) represents the repulsive electromagnetic vortex-vortex interaction with an effective London penetration depth $\lambda'$. The second term represents the attractive interaction of vortex lines due to the condensation energy of overlapping vortices with an effective coherence length $\xi'$. The vortex self energy or line tension is included in the diagonal terms $i=j$. A schematic sketch of the interactions is given in Fig.\,\,\ref{Vl_elast_struc_2}, panel (i).

\begin{figure}
\begin{center}
\includegraphics[width=0.45\textwidth]{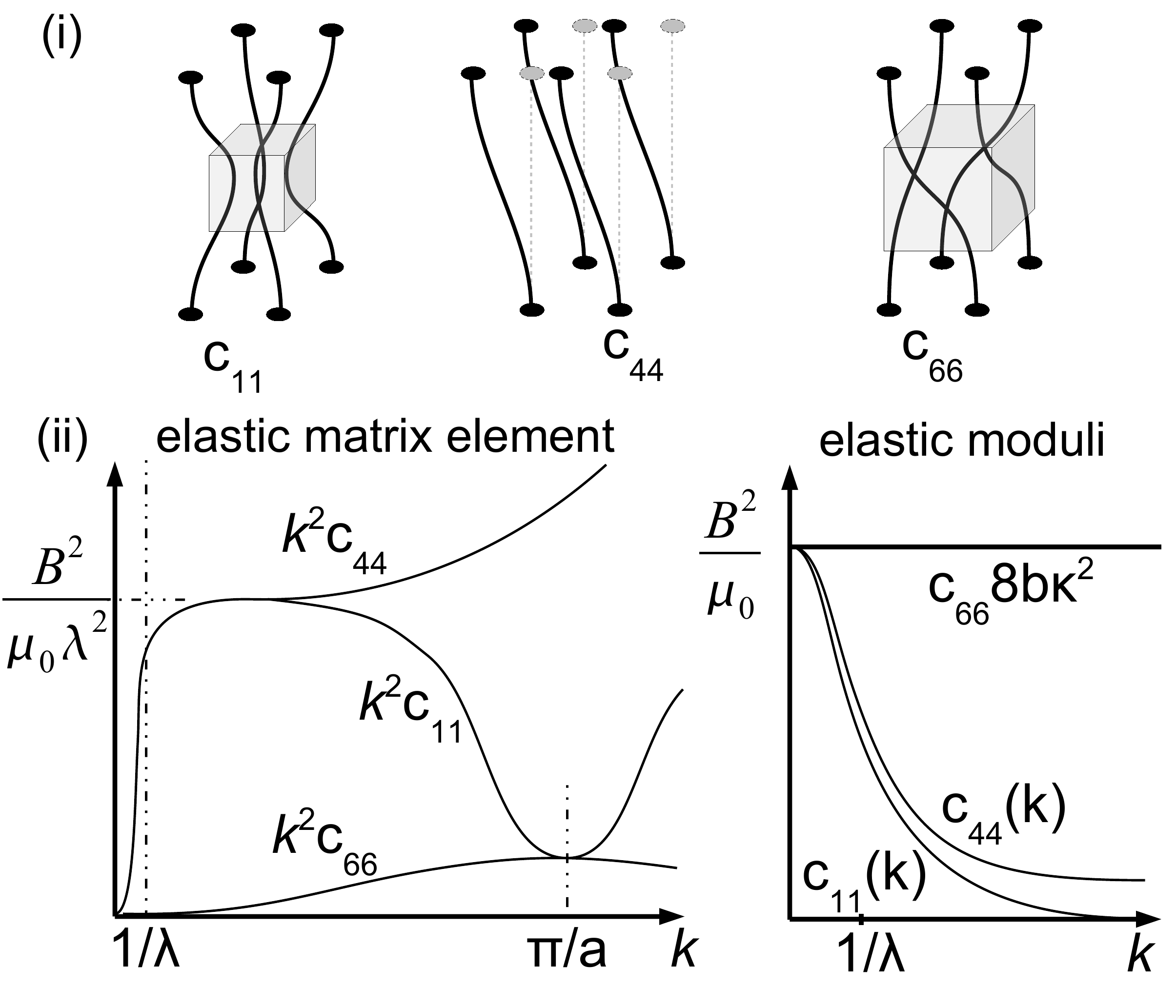}
\caption{ (i) Schematic depiction of the elastic constants $c_{11}$ for compression, $c_{44}$ for tilt and $c_{66}$ for shear of a superconducting VL. (ii) Depiction of the ${\bf k}$ dependence of the elastic matrix $\Phi_{\alpha,\beta}$ and of the elastic constants $c_{11}$, $c_{44}$ and $c_{66}$ of the VL. Note the different scaling for $c_{66}$ \cite{Brandt:95}.}
\label{elastic_constants}
\end{center}
\end{figure}
The elastic energy of a distorted VL, caused by pinning, structural defects, field gradients or transport currents, temperature gradients or thermal fluctuations, is small for most cases. Therefore, it can be calculated by linear elastic theory expressed in $k$-space. The displacements of a vortex line ${\bf u}_i(z)={\bf r}_i(z)-{\bf R}_i(z)=(u_{i,x};u_{i,y};0)$ from its ideal position ${\bf R}_i =(X_i;Y_i;z)$ is expressed by its Fourier components
\begin{equation}
{\bf u}_i(z) = \int_{BZ} \frac{d^3k}{8\pi^3} {\bf u}({\bf k}) e^{i{\bf kR}_i}
\label{eq1}
\end{equation}
and
\begin{equation}
{\bf u}({\bf k})=\frac{\phi_0}{B} \sum_i \int dz \, \,{\bf u}_i(z)e^{-i{\bf kR}_i}.
\label{eq2}
\end{equation}
With ${\bf u}({\bf k})=(u_x;u_y;0)$ the elastic free energy reads
\begin{equation}
F_{{\rm elast}}=\frac{1}{2} \int_{BZ} \frac{d{\bf k}}{8\pi^3} u_{\alpha}({\bf k}) \Phi_{\alpha \beta}({\bf k}) u_{\beta}^\star(\bf k)
\end{equation}
where $(\alpha,\beta) =(x,y)$. The integrals in eq. (\ref{eq1}) and eq. (\ref{eq2}) cover the first Brillouin zone of the VL in $k$-space and $-\xi^{-1} \leq k_z \leq \xi^{-1}$, respectively. $\Phi_{\alpha \beta}({\bf k})$ is called the elastic matrix of the VL. 

$\Phi_{\alpha \beta}({\bf k})$ is real, symmetric and periodic in $k$-space and is related to the elastic moduli $c_{11}$ for compression, $c_{44}$ for tilt and $c_{66}$ for shear within continuum theory by
\begin{equation}
\begin{split}
& \Phi_{\alpha \beta}({\bf k}) =(c_{11} - c_{66})k_{\alpha}k_{\beta} + \\
& \delta_{\alpha \beta}[k_{\perp}^2 c_{66} + k_z^2 c_{44} + \alpha_L({\bf k})] 
\end{split}
\end{equation}
with $k_{\perp}^2=(k_x^2 +k_y^2)$ (cf. Fig\,\,\ref{elastic_constants}). The $k$-dependence of the elastic matrix is plotted in Fig.\,\,\ref{elastic_constants}, panel (ii). The Labusch parameter  $\alpha_L$ describes the elastic interaction of the VL with pinning potentials caused by material inhomogeneities. For individual pinning, $\alpha_L$ is $k$-independent \cite{Labusch:69}, for weak collective pinning \cite{Larkin:79}, $\alpha_L({\bf k})$ decreases when $k_{\perp} >R_c^{-1}$ or $k_z > L_c^{-1}$ where $R_c$ and $L_c =(c_{44}/c_{66})^{1/2} R_c$ are the radius and length of the coherent short range ordered regions of the pinned VL.

\subsection{Uniform Distortions}
For uniform distortions, the elastic moduli of the VL may be written as \cite{Brandt:95}
\begin{equation}
\begin{split}
c_{11}& - c_{66} = \frac{B^2 \partial^2 F}{\partial B^2} = \frac{B^2 \partial \mu_0 H}{\mu_0 \partial B}\\
c_{44} &= \frac{B \partial F}{\partial B} = \frac{B \mu_0 H}{\mu_0} = B H\\
c_{66} &\approx \left(\frac{B \phi_o}{16 \pi \lambda^2 \mu_0} \right) \left( 1-\frac{1}{2\kappa^2} \right) (1-b)^2
\end{split}
\label{elasticity}
\end{equation}
with the Ginzburg-Landau parameter $\kappa$. $c_{11} -c_{66}$ is the modulus for isotropic compression. $H$ is the applied field, which is in equilibrium with the VL at the equilibrium induction $B$, given by the magnetization $M=\mu_0 H - B \leq 0$. The response of a VL to a change of the magnetic field direction is characterized by the tilt modulus $c_{44}$ of the VL. Note that $c_{66}$ vanishes either for $B\rightarrow B_{c2}$, which corresponds to strongly overlapping vortex cores, for $\lambda \rightarrow \infty$, corresponding to strongly overlapping vortex fields, or for $\kappa = 1/\sqrt{2}$. In the special case $\kappa = 1/\sqrt{2}$, all VL arrangements have the same free energy \cite{Brandt:95}.

Due to the long effective interaction lengths $\lambda'$ and $\xi'$, $c_{11}$ and $c_{44}$ strongly depend on the $k$-vector of the disturbance which is referred to as non-locality of the VL. This leads to a strong softening of the VL for short range distortions. This non-locality gives rise to large distortions, caused by pinning, disorder or thermal fluctuations, whereby the VL reacts to external forces in the form of a sharp cusp and not like pulling a string (cf. Fig.\,\,\ref{Vl_elast_struc_2}).

\subsection{Characteristic Timescales}
The elastic matrix of superconducting VLs determines the restoring force of ${\bf k}$-dependent distortions. The eigenmodes and the characteristic damping of VL fluctuations is determined by the restoring force as well as the viscous damping of the VL motion. The movement of the vortices with the velocity $v$ is damped by the viscosity 
\begin{equation}
\eta = \frac{B^2}{\rho_{\rm FF}}\approx \frac{B B_{c2}}{\rho_n} .
\label{damping}
\end{equation}
 This creates a drag force $v \eta$ per unit volume, where $\rho_{\rm FF}$ represents the flux-flow and $\rho_n$ the normal conducting resistivity. The elastic eigenmodes of the VL are given by a diagonalization of the elastic matrix $\Phi_{\alpha \beta}$. The result is a compressional and a shear eigenmode, relaxing with exponential time dependencies \cite{Brandt:95}. In continuum approximation, this yields:

\begin{equation}
\begin{split}
\Gamma_1({\bf k})& = (c_{11}({\bf k})k_{\perp}^2 + c_{44}({\bf k})k_z^2)/\eta \approx \Gamma_1 \\ 
\Gamma_2({\bf k})& = (c_{66}k_{\perp}^2 + c_{44}({\bf k})k_z^2)/\eta \approx \Gamma_1 k_z^2/k^2
\end{split}
\end{equation} 
For the VL in a typical clean low $\kappa$ superconductor as e.g. Nb, the the characteristic relaxation rates $\Gamma_1$ and $\Gamma_2$ of the VL are in the range of 10$^{-9}$ s$^{-1}$.

\section{Experimental Setup}
\label{Experimental_setup}

To measure the VL tilt modulus $c_{44}$ by means of time resolved SANS, a time varying magnetic field setup, consisting of two orthogonal pairs of Helmholtz-coils has been designed. By imposing a periodic tilting of the magnetic field, we induce a relaxation process of the VL which can be described by a diffusion process. The relaxation of the VL is measured by means of time resolved stroboscopic SANS. The measurements have been performed on the small angle diffractometer V4 at BENSC \cite{V4, Wiedenmann:06}. We introduce the details of the experimental setup used for our study in the following paragraphs:

\subsection{Sample Environment and Magnetic Field Setup}

\begin{figure}[h]
\begin{center}
\includegraphics[width=0.37\textwidth]{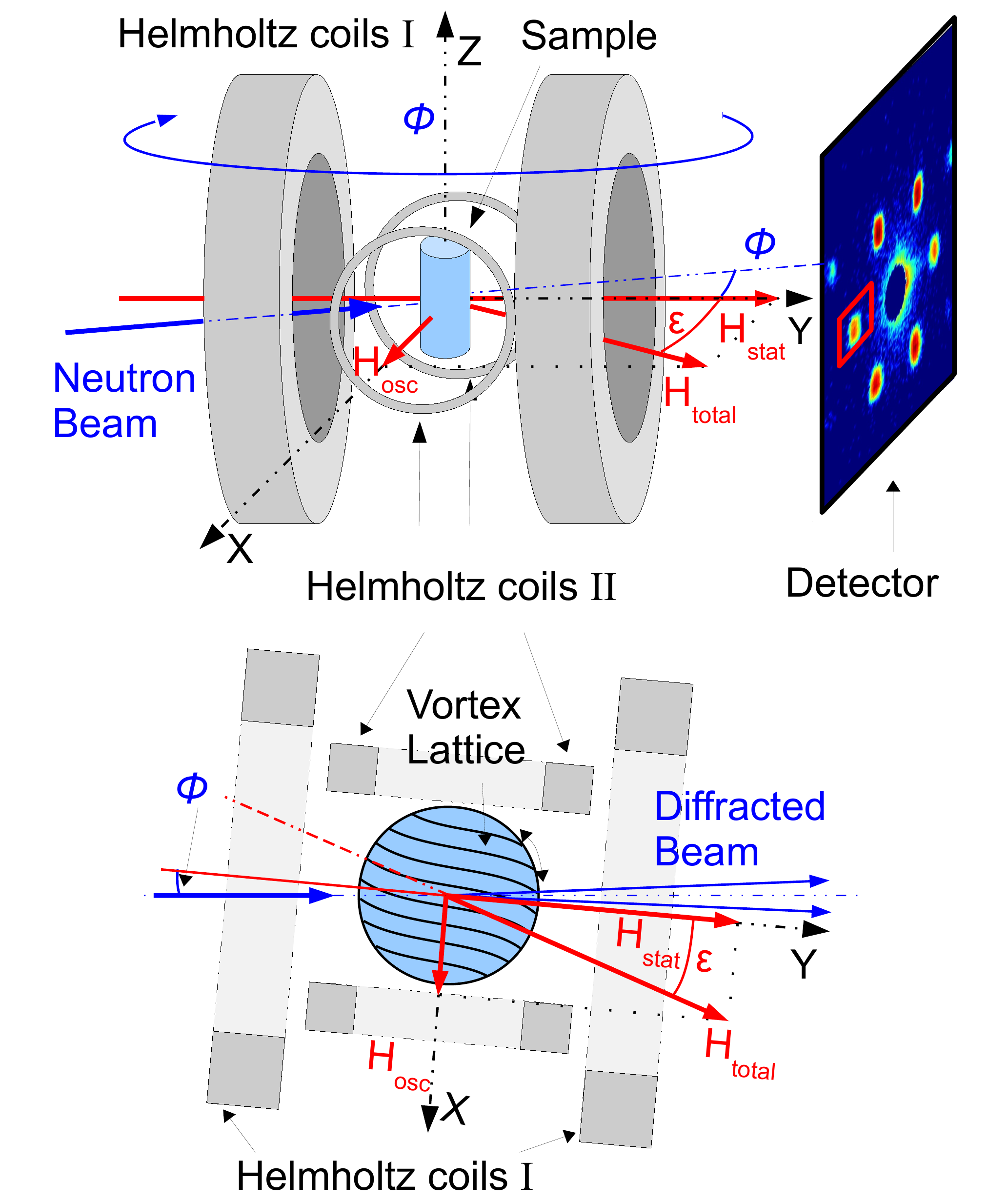}
\caption{Schematic depiction of the experimental setup used for time resolved SANS measurements of the VL tilt modulus $c_{44}$. The sample is located in the centre of two orthogonal magnetic fields, generated by Helmholtz coils. With a combination of a static magnetic field $H_{\rm stat} \parallel y$ and a time-varying magnetic field $ H_{\rm osc} \parallel x $ with $H_{\rm osc} \ll H_{ stat}$, the resulting magnetic field can be rotated with respect to the sample in the $xy$ plane. Both magnetic fields and the sample can be rocked around the vertical $z$-axis with respect to the incoming neutron beam by the angle $\phi$. The resulting magnetic field $H_{total}$ is roughly parallel to the incoming neutron beam. The scattered intensity is recorded on a two-dimensional detector.}
\label{setup}
\end{center}
\end{figure}

A schematic drawing of the experimental setup is given in Fig.\,\,\ref{setup}: The sample is located in the center of two Helmholtz-coils, cooled with a closed-cycle cryostat to a minimum temperature of 4\,K. Both magnetic fields and the sample can be rocked together with respect to the vertical z-axis. The static main field ${\bf H}_{{\rm stat}}$, applied along the y-axis is generated by bespoke water cooled copper coils \cite{Muehl:05}. ${\bf H}_{\rm stat}$ is oriented approximately parallel to the incoming neutron beam. The angle enclosed between ${\bf H}_{\rm stat}$ and the incoming neutron beam, which is also applied in the xy plane, is denoted rocking angle $\phi$. A magnetic field 75\,mT$\leq {\bf H}_{\rm stat} \leq $135\,mT was applied.

The time varying field ${\bf H}_{{\rm osc}}$ is generated by a small air-cooled set of Helmholtz-coils inside the main coil, driven with an arbitrary waveform generator and an amplifier. ${\bf H}_{{\rm osc}}$ is oriented along the x-axis perpendicular to ${\bf H}_{{\rm stat}}$. A rectangular pulse shape with an amplitude of ${\bf H}_{\rm osc}=0\,\rm mT\leftrightarrow 5\,\rm mT$ and a repetition rate of 0.2\,Hz was applied. The resulting field ${\bf H}_{{\rm total}}$ is rotated with respect to ${\bf H}_{{\rm stat}}$ by the angle $\epsilon= \arctan{|{\bf H}_{\rm osc}|/|{\bf H}_{\rm stat}|}$ in the xy plane. ${\bf H}_{{\rm stat}} \gg {\bf H}_{{\rm osc}}$ yields that $|{\bf H}_{{\rm total}}| \approx |{\bf H}_{{\rm stat}}| = |{\bf H}|$. Due to the perpendicular alignment we omit the vectorial notation of ${\bf H}_{{\rm stat}}$ and ${\bf H}_{{\rm osc}}$. A smearing of the applied pulses is caused by the rise and fall time of amplifier used for $H_{{\rm osc}}$. It has been determined with a Hall-probe at the sample position and found to be $\sim$ 5\,ms.

Hence two equilibrium positions for the magnetic field and also the VL emerge, which are separated by $\epsilon \approx 2^{\circ}$ if $H_{\rm osc}$ is alternated between $H_{\rm osc}=$0\,mT and $H_{\rm osc}=$5\,mT. The cylindric sample is aligned with its symmetry axis parallel to the z-axis, i.e. a constant demagnetizing factor $N = 1/2$ applies for all angles $\epsilon$.

For the magnetic field range of our experiment the VL assumes a six-fold scattering pattern with a Bragg angle of a few tenths of a degree. The instrumental resolution in the direction perpendicular to the scattering vector $\bf q$ yields a value of 0.2$^{\circ}$ for a collimation length of $L_1=8\,\rm{m}$, a sample detector distance of $L_2=8\,\rm{m}$, a source aperture of $R_1=10\,\rm{mm}$ and a sample aperture of $R_2=2\,\rm{mm}$, respectively \cite{Pedersen:90}. The shift of the VL direction of $\epsilon \approx 2^{\circ}$ is hence much larger than the instrumental resolution giving rise to a large contrast.

\subsection{Measurement Principle}
In the following we describe the principle used for the measurement of the VL motion and relaxation, driven by the time dependent transverse magnetic field. The rocking angle $\phi$ is initially adjusted to satisfy the Bragg condition for a reciprocal lattice vector of the VL ${\bf q}={\bf G_{VL}}={\bf k}_i -{\bf k}_f$ (lying in the xy-plane) for $H_{\rm osc}=0\,\rm mT$, i.e. $\epsilon =0^{\circ}$. The observed scattering intensity at the 2D detector at ${\bf G_{VL}}$ thus is a measure for the quantity of VL which points in this direction. $H_{\rm osc}$ oscillates between $\mu_0 H_{\rm osc}=0$\,mT and $\mu_0 H_{\rm osc}=5$\,mT. Thus, the relaxation process between these two equilibrium positions can be followed by measuring the integrated intensity at the Bragg reflection at ${\bf G_{VL}}$ as function of time. Two different time-dependent processes can be measured. 

\begin{itemize}
\item The first process ($\mu_0 H_{\rm osc}$ decreases from 5\,mT to 0\,mT) is attributed to the VL relaxing {\it into} the Bragg condition \footnote{Note, that the instrumental resolution $\Delta \beta_{\bf kf}=0.2^{\circ}$ is significantly smaller compared to the angular separation of both VL equilibrium positions $\epsilon \approx 2^{\circ}$}. The corresponding time-scale is denoted $\tau_1$. 
\item The latter process ($\mu_0 H_{\rm osc}$ increases from 0\,mT to 5\,mT) describes a time-scale, necessary to pull the VL out of the Bragg condition denoted as $\tau_3$.
\end{itemize}

\begin{figure}[h]
\begin{center}
\includegraphics[width=0.34\textwidth]{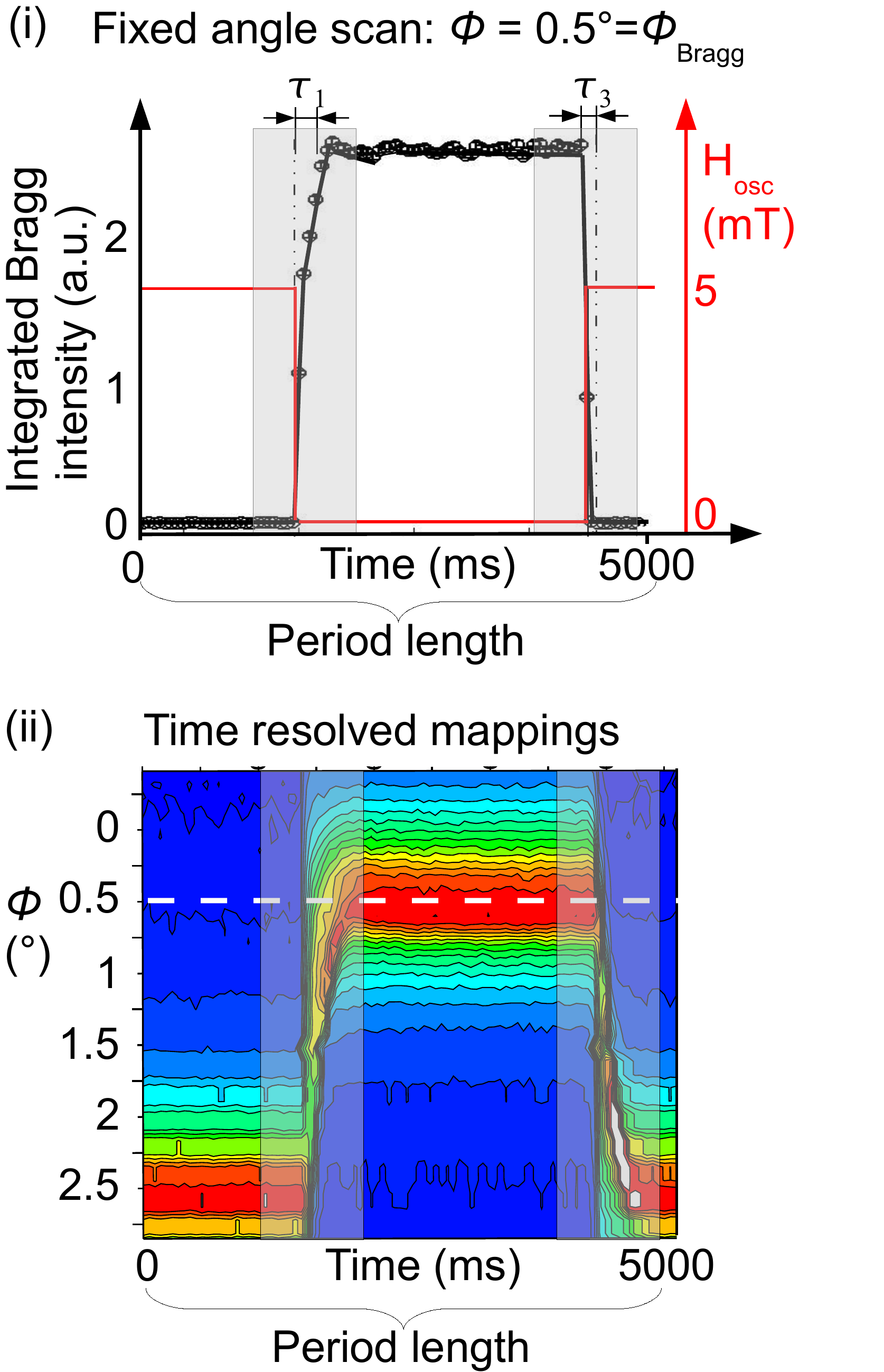}
\caption{Panel (i) shows a typical {\it fixed angle scan}. A transverse magnetic field deflects the direction of the VL in the sample leading to a drop in the Bragg intensity. The red line indicates the modulus of $H_{\rm osc}$. Panel (ii) shows a typical time resolved map for varying rocking angles $\phi$. Two equilibrium positions for the VL, as induced by the change of $H_{\rm osc}$, are visible for $\phi = 0.5^{\circ}$ and $\phi = 2.5^{\circ}$. The switching process between these equilibrium positions can be monitored as a function of time. The horizontal white line in panel (ii) represents the {\it fixed angle scan} given in panel (i). Both, scans (i) and (ii) have been performed at $T=4\,{\rm K}$ and $\mu_0H=$100\,mT.}
\label{setup_2}
\end{center}
\end{figure}

In the following, this method is denoted {\it fixed angle scan}. Typical data for a temperature of $T=4\,{\rm K}$ and a magnetic field of $\mu_0H=$100\,mT is shown in Fig.\,\,\ref{setup_2}, panel (i), where the red line indicates the modulus of $H_{\rm osc}$. The whole relaxation process of the VL can be traced angle and time resolved, when such scans are performed for each rocking angle $\phi$. The latter method is denoted {\it time resolved mapping}. A representative scan is given in Fig.\,\,\ref{setup_2}, panel (ii) for $T=4\,{\rm K}$ and $\mu_0H=$\,100mT. {\it Fixed angle scans} are represented by cuts at a fixed rocking angle $\phi$ in the {\it time resolved mappings}. It is important to note that the angular distribution of the VL is always integrated over the complete sample and thus is additionally convoluted with the intrinsic VL mosaicity and the angular resolution of the small angle scattering instrument.

\subsection{Stroboscopic Small Angle Neutron Scattering}

The fundamental principle of stroboscopic neutron scattering is the excitation of the sample by an external control parameter followed by a measurement of the time dependent dynamic response and relaxation \footnote{Note, that only elastic scattering is considered for the stroboscopic small angle neutron scattering technique used for our experiment.}, in our case the direction of the magnetic field. To increase time resolution and signal statistics, these measurements are performed in a {\it stroboscopic} manner, i.e. the measurement is repeated many times where the data obtained for the individual cycles is summed coherently \cite{Wiedenmann:06}. The stroboscopic SANS technique is realized, using a standard SANS setup, extended by a {\it time resolved} position sensitive detector. The repetition cycles of the time resolved detector and the control parameter are phase locked.  

The time resolution is mostly determined by a smearing of the single frames, caused by the wavelength spread $\Delta \lambda/\lambda$ of the neutron beam leading to a variation of the neutron time of flight from the sample to the detector. The time of flight is given by the equation \cite{Wiedenmann:06} $t_{\rm TOF}[{\rm ms}]=\lambda [{\rm \AA}] \cdot L_2 [{\rm m}] \cdot 0.253$. $\lambda$ represents the wavelength of the neutron beam and $L_2$ the sample-detector distance, respectively. For SANS measurements, $L_2$ and $\lambda$ determine the accessible $q$-range. A large sample-detector distance $L_2$ and large wavelength $\lambda$ is desirable to resolve large real space structures associated with small $q$-vectors. This leads to a significant loss in time resolution. For the measurements on the VL in Nb, presented in this manuscript, a wavelength $\lambda=$8\AA\,, a wavelength spread $\Delta \lambda/ \lambda=0.1$ and a detector distance $L_2$=8\,m lead to a time resolution of $\pm$1.6\,ms.

\subsection{Experimental Parameters}

\begin{figure}[h]
\begin{center}
\includegraphics[width=0.33\textwidth]{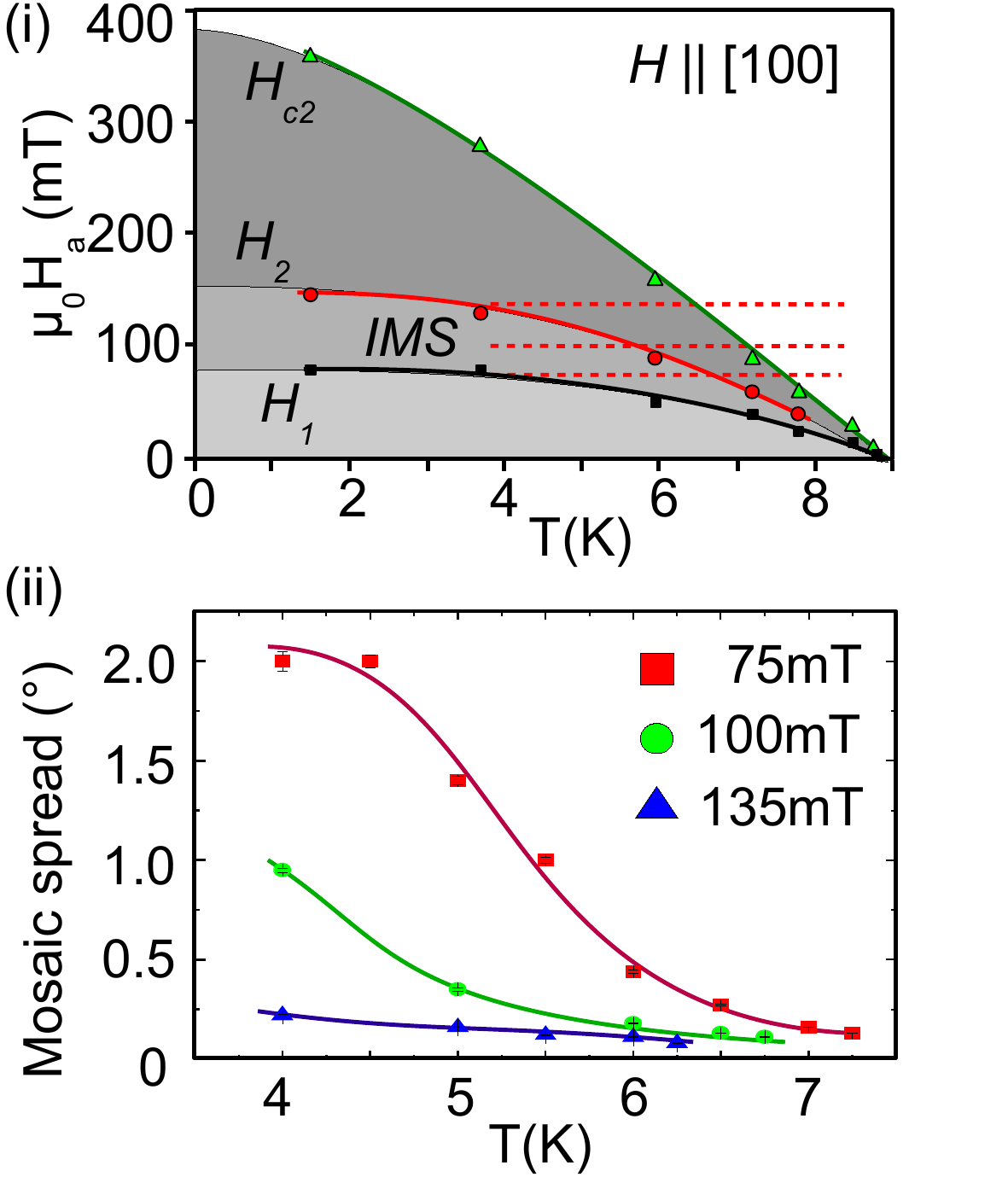}
\caption{Panel (i) depicts the phase diagram obtained by previous SANS measurements \cite{Muehlbauer:09}. A demagnetizing factor $N=1/2$ applies. The dashed red lines locate scans, performed for our measurements of $c_{44}$, in the phase diagram. Panel (ii) depicts the intrinsic mosaicity of the VL as measured with SANS as function of temperature for  magnetic fields $\mu_0 H=$ 75\,mT, 100\,mT and 135\,mT. The lines serve as guide to the eye.}
\label{mosaicity}
\end{center}
\end{figure}

Measurements of the tilt modulus $c_{44}$ of the VL have been performed for an applied magnetic field $\mu_0H=75\,\rm{mT}$, 100\,mT and 135\,mT, each for sample temperatures between 4\,K and $T_c$. A schematic sketch of the measurement range with respect to the phase diagram, obtained in previous SANS measurements \cite{Muehlbauer:09} is given in Fig.\,\,\ref{mosaicity}, panel (i): Both, the IMS and the crossover to the Shubnikov phase are covered, where the vortex-vortex interaction changes from attractive to repulsive. 

The intrinsic mosaicity of the VL as inferred from static rocking scans is given in panel (ii) of Fig\,\,\ref{mosaicity}: For the highest temperatures the mosaicity of the VL is limited by the instrumental resolution. For decreasing temperatures $T\leq6\,\rm K$ for $\mu_0H=75\,\rm {mT}$ and $T\leq5\,\rm{K}$ for $\mu_0H=100\,\rm {mT}$ the mosaic spread shows a pronounced increase. However, in the Shubnikov phase for $\mu_0H=$135\,mT only a weak increase of the mosaic spread is observed for decreasing temperatures. The increase of mosaicity for decreasing temperature is attributed to the crossover to the IMS where an additional bending of vortices is caused by the complicated VL domain structure in combination with increasing Meissner effect.

To avoid hysteretic effects, all measurements have been taken after cooling in a field to the desired temperature (FC). Furthermore, the direction of the magnetic field is oscillating continuously due to $H_{\rm osc}$. This leads to an effective depinning of the VL. An equilibrium state can thus be assumed. In analogy to HFC paths in the B-T phase diagram of type-I superconductors in the intermediate state, where the magnetic flux is expelled for decreasing field, a similar behaviour and morphology is expected in the IMS for type-II superconductors upon FC. The result is supposed to be an open, multiply connected topology of Shubnikov domains enclosing regions of Meissner phase. In addition, the IMS is characterized by Landau branching of the Shubnikov-domains at the surface of the sample.

\subsection{Sample Used for our Study}
\label{sample}
For our studies, a cylindrical Nb single crystal with a length of 20\,mm and a diameter of 4.5\,mm was cut by spark erosion from a rod that had been produced at the ZFW Dresden more than 30 years ago. The cylindric symmetry axis of the sample coincides with a crystallographic $\langle$110$\rangle$ axis. A further $\langle$110$\rangle$ axis is oriented parallel to the incident neutron beam. The preparation process of the rod consisted of purification by liquid-liquid extraction combined with chlorination and thermal decomposition of NbCl$_5$ followed by electron beam floating zone melting, decarburization in oxygen atmosphere and annealing in UHV \cite{Berthel:76, Koethe:00}. The impurity content was estimated to be less than 1\,ppm for interstitial and better than 2\,ppm for substitutional impurities. 

To remove the surface layer of the cutted sample and to decrease the surface roughness, the sample was etched with a mixture of concentrated HF and HNO$_3$ for several minutes. In order to remove interstitials, in particular hydrogen introduced due to the spark erosion cutting process as well as the long storage, the sample was again RF-annealed in UHV above $2000^{\circ}$ at the University of Birmingham for one week followed by surface oxygenation to reduce the Bean-Livingston barrier for surface pinning \cite{Laver:06}. The sample shows a smooth, highly polished surface with a pale golden colour without any signs of pores, scratches or damages.

The residual resistivity ratio (RRR) of the sample was measured with an eddy current decay method at the University of Birmingham. The RRR was extrapolated to $T\to 0$ at $B=0$ for a temperature range of 14.5\,K - 9.3\,K assuming $\rho_{phonon} \propto T^3$ as well as extrapolated to $B=0(T=4.2\,{\rm K})$ from $B_{c2}(T=4.2\,{\rm K})$ assuming $\rho_B \propto B$, yielding values from RRR$=$8000 to 16000, respectively. However, the temperature extrapolation is more reliable, leading to a value of RRR$\sim10^4$. This leads to a Ginzburg-Landau coefficient $\kappa \sim $ 0.74 at 0.9$T/T_c$ \cite{Berthel:76}. The AC susceptibility and the magnetization, measured at the Technische Universit\"at M\"unchen were consistent with the literature. Previous small angle neutron scattering measurements on the sample have shown no indications of trapped flux for decreasing magnetic field, further indicating an excellent sample quality \cite{Muehlbauer:09} and a vanishing surface pinning barrier for the vortices.

The flux-flow resistivity $\rho_{\rm FF},$ responsible for the damping of vortex motion in superconductors is related to the normal conducting resistivity $\rho_n$ by eq. (\ref{damping}). For high purity Nb in the clean limit $\rho_n$ was numerically approximated \cite{Berthel:76} by
\begin{equation}
\begin{split}
\rho_n(T)= &\rho_0+ c_n \left( bT^2 + \frac{cT^3}{7.212} \int_{\Theta_{\rm min} /T}^{\Theta_D/T} \frac{x^3}{(e^x -1)(1-e^x)} dx+ \right.\\
& \left.+ \frac{dT^5}{124.4} \int_{\Theta_{\rm min} /T}^{\Theta_D/T} \frac{x^5}{(e^x -1)(1-e^x)} dx \right) ,
\end{split}
\label{rho}
\end{equation}
with $b=1.63\cdot 10^{-6} {\rm K}^{-2}$, $c=(2.864 \pm 0.003)\cdot 10^{-7} {\rm K}^{-3}$, $d=(1.81 \pm 0.004)\cdot 10^{-10} {\rm K}^{-5}$ a Debye temperature of $\Theta_D=270$\,K, $\Theta_{\rm min}=35$\,K and the normalization constant $c_n = 1.8\cdot 10^{-7}$. The RRR of $10^4$ and the literature value for $\rho_n$(300\,K)$\sim 1.3\cdot 10^{-7}\,\Omega$m \cite{Webb:69} lead to $\rho_0\sim 1.3\cdot 10^{-11}\,\Omega$m. This yields an increase of the normal conduction resistivity $\rho_n$ in the relevant temperature range of the experiment from 4\,K to 8\,K by a factor of about 2 from 1.8\,n$\Omega$cm to 3.2\,n$\Omega$cm.

\section{Experimental Results}
\label{Results_stroboscopic}

\subsection{Time Resolved Mappings}
In the following we present our data obtained for the VL relaxation and diffusion. First, we focus on the {\it time resolved mappings}. Fig.\,\,\ref{bigpics_all} depicts the relaxation of the VL at a magnetic field $\mu_0 H$=100\,mT for a temperature of $T=6.5\,{\rm K}$ (panels (i) and (ii)) and $T=4\,{\rm K}$ (panels (iii) and (iv)). The integrated Bragg intensity is plotted on a linear scale. The horizontal broken white lines marked with the black arrows indicate the time when the magnetic field direction is switched between the two equilibrium positions. The time-range displayed corresponds to the gray shading in Fig.\,\,\ref{setup_2}, panel (ii), however, the axes have been rotated for better visibility. Both equilibrium positions (indicated with vertical broken white lines) at $\phi = 0.5^{\circ}$ and $\phi = 2.75^{\circ}$ are clearly visible for the data obtained at $T=4\,{\rm K}$. The measurement range was reduced for $T=6.5\,{\rm K}$ due to the limited beamtime. Salient features of the VL relaxation are:

\begin{figure}
\begin{center}
\includegraphics[width=0.5\textwidth]{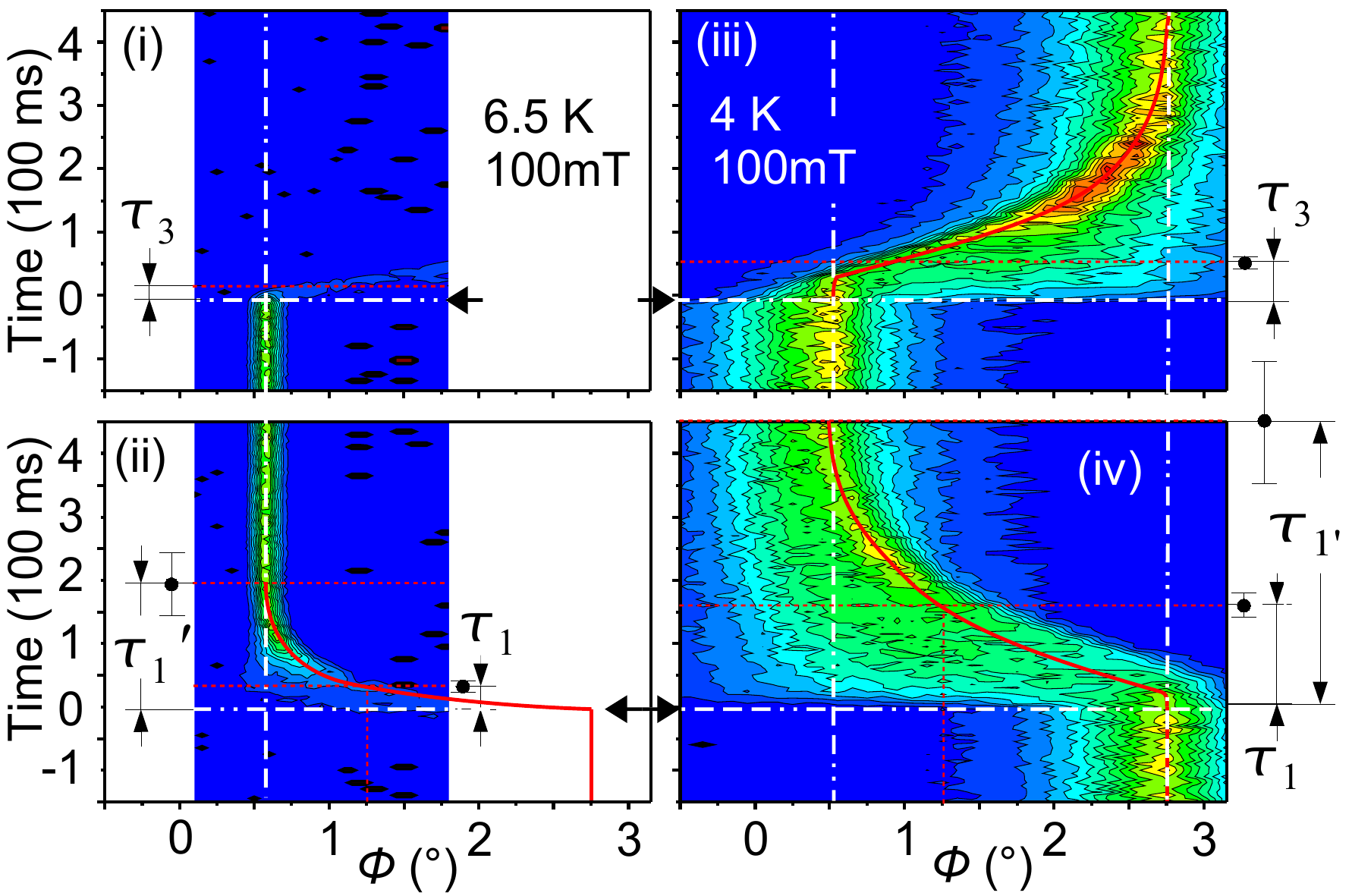}
\caption{Panels (i) and (ii) depict {\it time resolved mappings} for an applied magnetic field $\mu_0 H$=100\,mT and a temperature of 6.5\,K. Panels (iii) and (iv) depict similar scans for $\mu_0 H$=100\,mT and 4\,K. Note the reduced measurement range for $T=6.5\,{\rm K}$. The contours are plotted on a linear scale. The change of magnetic field direction is indicated by horizontal broken white lines marked with black arrows, whereas the equilibrium positions of the VL are marked with vertical broken white lines. The continuous red lines indicate the relaxation process of the VL. The red lines serve as guide to the eye.}
\label{bigpics_all}
\end{center}
\end{figure}

\begin{itemize}
\item A larger intrinsic mosaicity of the VL is observed for $T=4\,{\rm K}$ in comparison to $T=6.5\,{\rm K}$. The additional mosaic spread, caused by a possible bending of the vortices due to the domain structure of the IMS in combination with demagnetizing effects is in agreement with static data as displayed in panel (ii) of Fig.\,\,\ref{mosaicity}.

\item Relaxation time $\tau_1$: As expected, the general trend is a relaxation characteristic of an exponential decay $\propto e^{-t/\tau_1}$ for both $T=4\,{\rm K}$ and $T=6.5\,{\rm K}$, indicated by the continuous red lines. The corresponding time constant $\tau_1$, obtained for $T=6.5\,{\rm K}$ yields $\tau_1=0.04\pm\,0.008\,\rm s$, whereas $\tau_1=0.16\pm\,0.02\,\rm s$ for $T=4\,{\rm K}$. In addition, we further define $\tau_1'$ as the characteristic time-scale, when the VL has reached its new equilibrium position. For $T=6.5\,{\rm K}$ yields $\tau_1'=0.2\pm\,0.05\,\rm s$, whereas $\tau_1'=0.45\pm\,0.1\,\rm s$ for $T=4\,{\rm K}$. 

\item Relaxation time $\tau_3$: $\tau_3$ indicates the time, at which the intensity at the Bragg spot ${\bf G}_{VL}$ has decreased to $1/e$ of its initial value. We obtain $\tau_3=0.06\pm0.008\,\rm s$ for $T=4\,{\rm K}$ and $\tau_3=0.025\pm0.005\,\rm s$ for $T=6.5\,{\rm K}$, respectively. 

\item The intensity map, obtained for $T=4\,{\rm K}$ is characterized by a drastic increase of mosaic spread immediately after the magnetic field direction is changed. The time-scale observed for this feature is found to be less than $\tau_3$. In contrast, only a moderate increase of mosaic spread is observed for $T=6.5\,{\rm K}$.
\end{itemize}

\subsection{Fixed Angle Scans}
\begin{figure}[h]
\begin{center}
\includegraphics[width=0.5\textwidth]{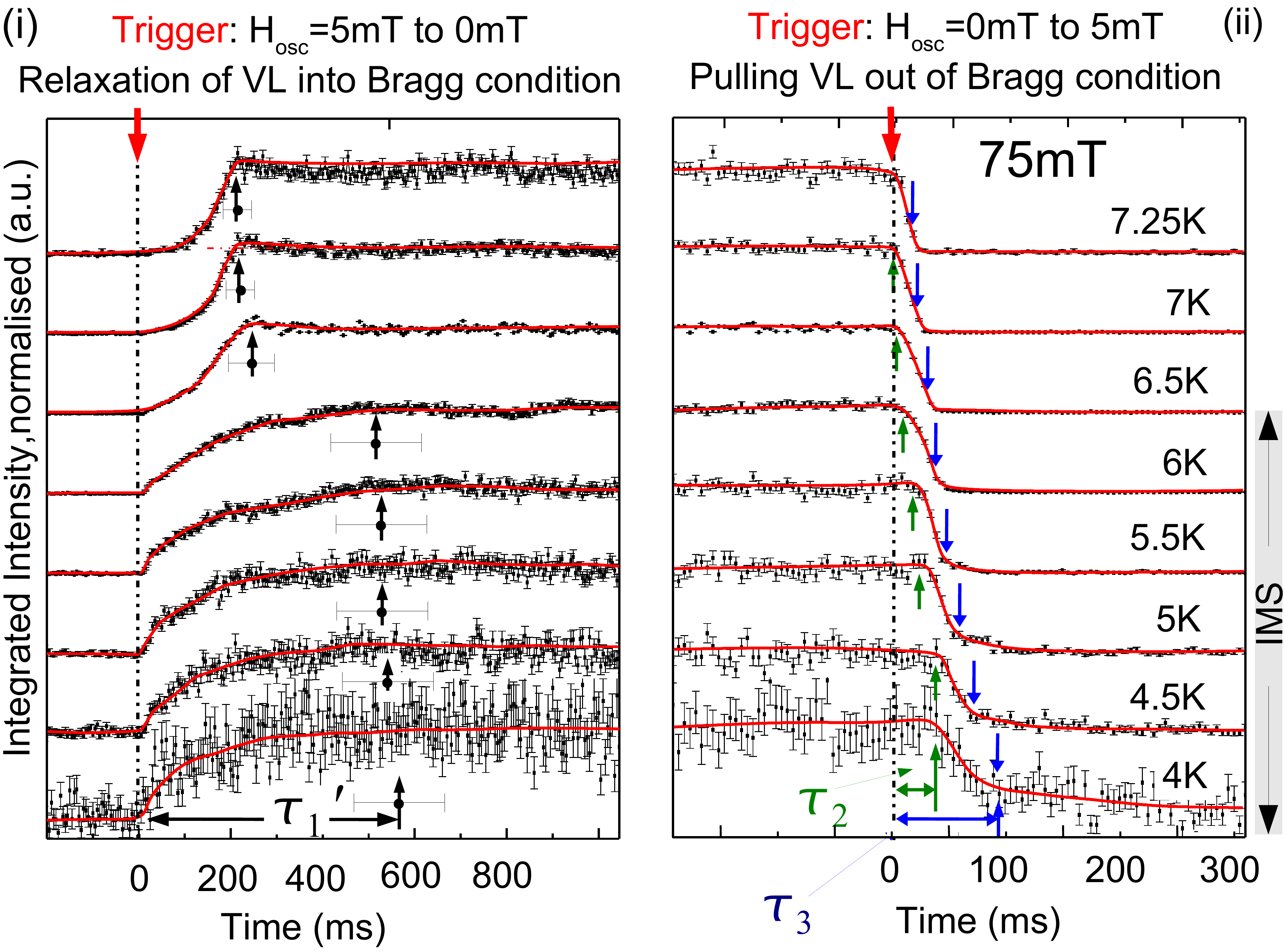}
\caption{Panel (i) shows {\it fixed angle scans} for $\mu_0H=$75\,mT for increasing temperature as labeled in panel (ii). The increase of scattering intensity after the trigger is attributed to the relaxation of the VL {\it into} the Bragg condition as function of time. The characteristic relaxation time is denoted $\tau_1'$, indicated with black markers. Panel (ii) shows {\it fixed angle scans} for $\mu_0H=$75\,mT  for increasing temperature. The decrease of intensity after the trigger is attributed to the VL being pulled {\it out} of the Bragg condition. The characteristic relaxation times are denoted $\tau_2$ and $\tau_3$, indicated with green and blue markers, respectively. Note the different timescale for panel (i) compared to panel (ii). The different plots have been shifted vertically for clarity reasons, the lines serve as guide to the eye. The phase region of the IMS in indicated on the rhs.}
\label{tscans_1}
\end{center}
\end{figure}

New light is shed on the details of the VL relaxation process, focusing on the {\it fixed angle scans} as function of applied magnetic field $H$ and temperature $T$. The data is given in Fig.\,\,\ref{tscans_1}, panels (i) and (ii) for a magnetic field of $\mu_0H=75\,\rm mT$,  Fig.\,\,\ref{tscans_2}, panels (i) and (ii) for $\mu_0H=100\,\rm mT$ and Fig.\,\,\ref{tscans_3}, panels (i) and (ii) for $\mu_0H=135\,\rm mT$. The time-range displayed for the three figures corresponds to the gray shadings in Fig.\,\,\ref{setup_2}, panel (i).

\begin{figure}
\begin{center}
\includegraphics[width=0.45\textwidth]{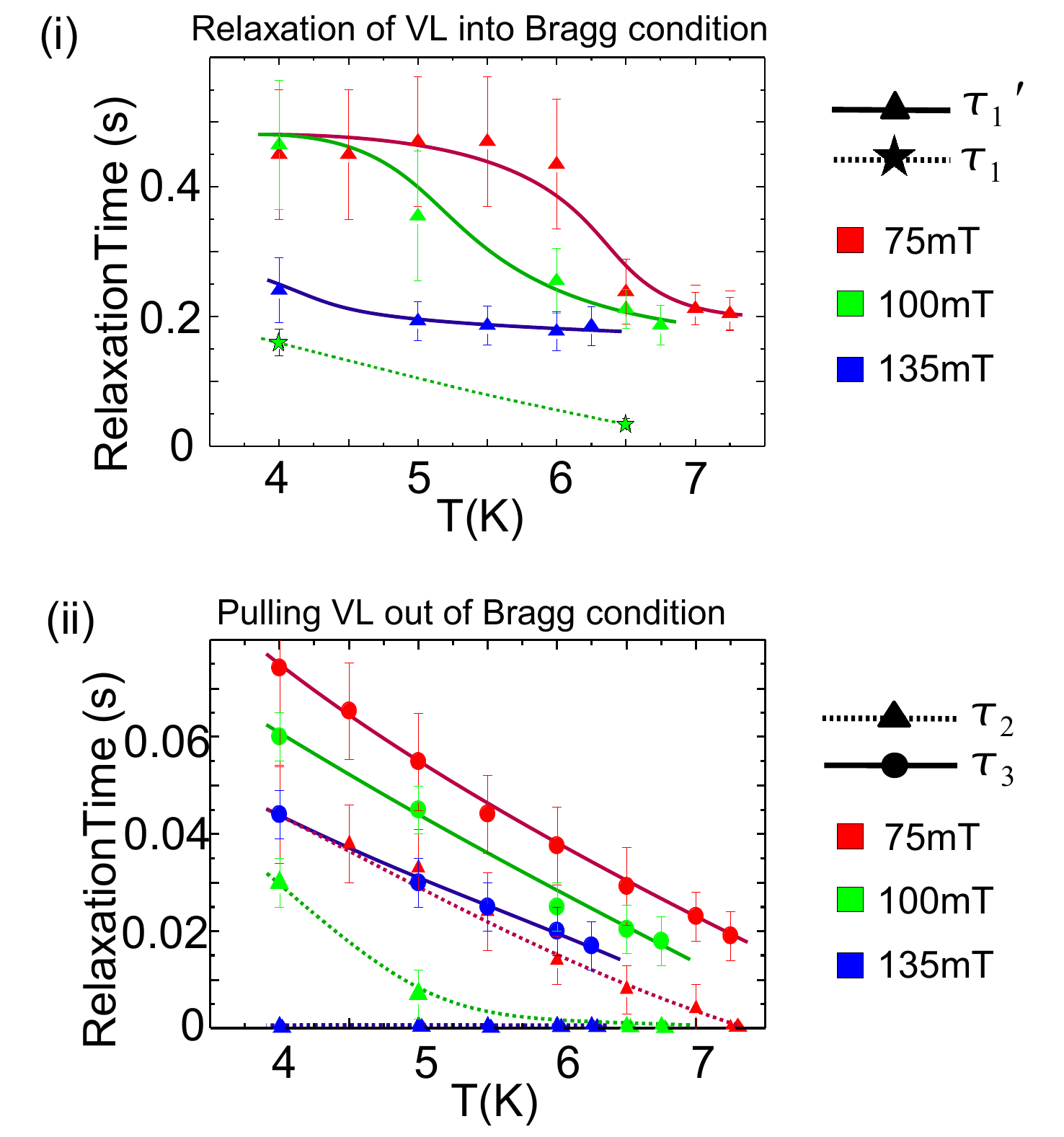}
\caption{Panels (i) and (ii) show the  values obtained for time constants $\tau_1$, $\tau_1'$, $\tau_2$ and $\tau_3$ as function of temperature $T$ and magnetic field $\mu_0H$. Note the different scaling for panel (i) and panel (ii). The lines serve as guide to the eye.}
\label{diffusionconstants_3}
\end{center}
\end{figure}

We first focus on panels (i) of Figs.\,\,\ref{tscans_1}, \ref{tscans_2} and \ref{tscans_3} which show {\it fixed angle scans} for increasing sample temperature from $T=4\,{\rm K}$ to $T_c$ (as labeled in panels (ii)). The increase of scattering intensity after the vertical line, labeled with {\it trigger} is attributed to the relaxation of the VL {\it into} the Bragg condition, the characteristic relaxation time was denoted $\tau_1$. However, it turns out that the qualitative shape of the relaxation process changes as a function of temperature. It is therefore not possible to determine the characteristic time-scale $\tau_1$, in a way similar to the {\it time resolved mappings} \footnote{Due to the limited beamtime, the systematic temperature and magnetic field dependence of the VL relaxation was measured only for the {\it fixed angle scans}.}. In order to cover the systematic trends, the magnetic field and temperature dependence of $\tau_1'$ was analyzed. Salient features are:

\begin{itemize}
\item The same general trend as observed for the {\it time resolved mappings} can also be identified: Increasing temperature and increasing magnetic field lead to a significantly faster relaxation. The relaxation time $\tau_1'$ is indicated with a black marker in each {\it fixed angle scan}. The resulting magnetic field and temperature dependence of $\tau_1'$ is shown in Fig.\,\,\ref{diffusionconstants_3}, panel (i): For low temperatures $T\approx4\,\rm K$ and magnetic fields of $\mu_0H=75\,\rm {mT}$ and 100\,mT, $\tau_1'=$0.5$\pm$0.1\,s. For increasing temperatures, $\tau_1'$ decreases characteristic of a smooth crossover to values of $\tau_1'=$0.2$\pm$0.05\,s. The crossover temperature thereby decreases with increasing field from $T=6.5\,{\rm K}$ at $\mu_0H$=75\,mT to $T=5.5\,{\rm K}$ at $\mu_0H=100\,\rm mT$. In contrast, for a magnetic field of $\mu_0H=135\,\rm mT$, $\tau_1'$ assumes a constant value of $\tau_1=0.2\pm0.05\,\rm s$ except at the lowest temperature where a slight shift to $\tau_1'=0.25\pm0.05\,\rm s$ is observed. The crossover temperature for $\mu_0H=135\,\rm mT$ is expected to be at $\approx$ 3.5\,K, i.e. slightly below the accessible temperature regime. The magnetic field dependence of the crossover temperature of $\tau_1'$ corresponds to the crossover from the IMS to the Shubnikov phase. The data points obtained for $\tau_1$ from the {\it time resolved mappings} are given in Fig.\,\,\ref{diffusionconstants_3}, panel (i) as well.

\begin{figure}[h]
\begin{center}
\includegraphics[width=0.5\textwidth]{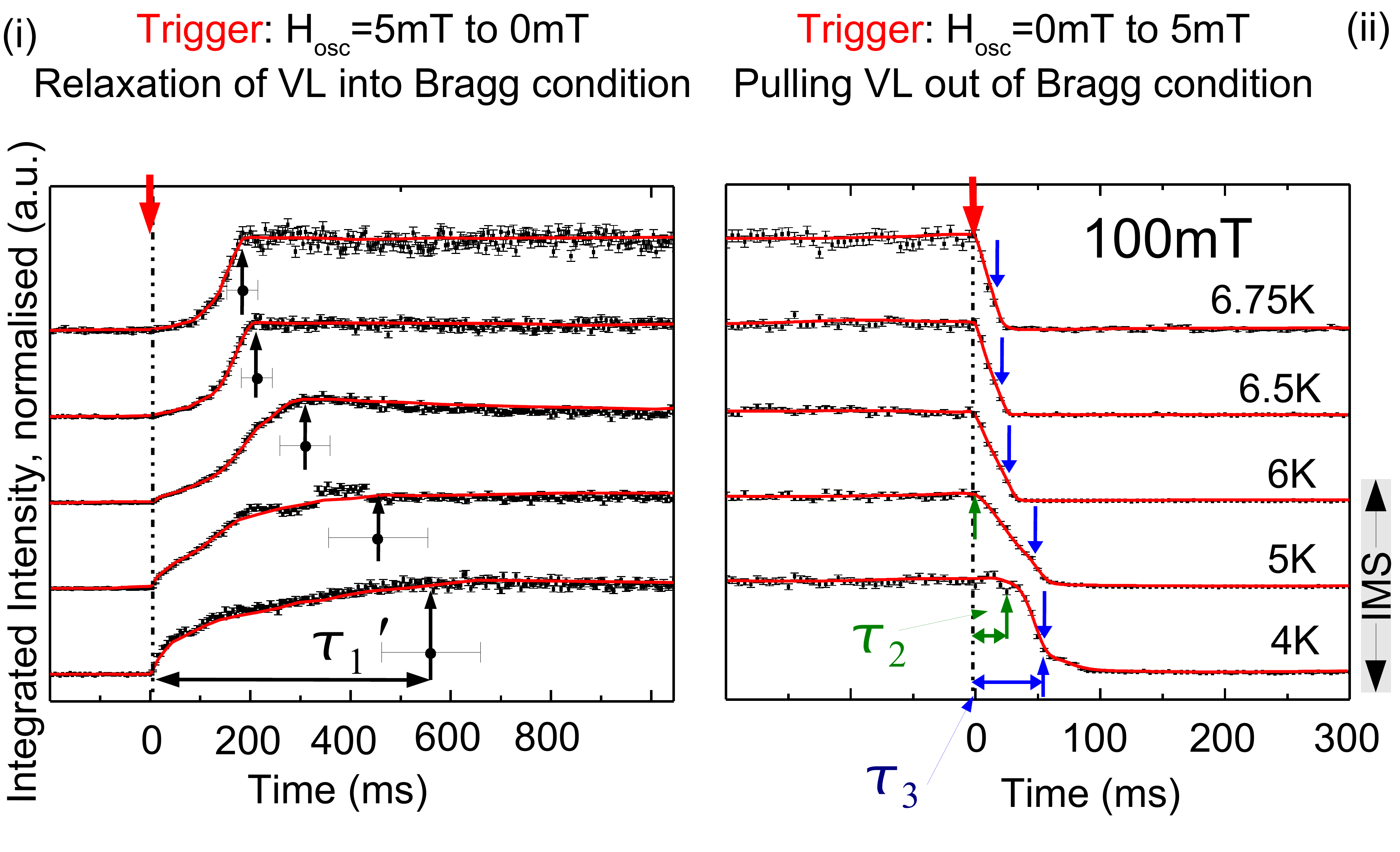}
\caption{Panel (i) shows {\it fixed angle scans} for $\mu_0H=$100\,mT analogous to Fig.\,\,\ref{tscans_1}.}
\label{tscans_2}

\vspace{0.5cm}

\includegraphics[width=0.5\textwidth]{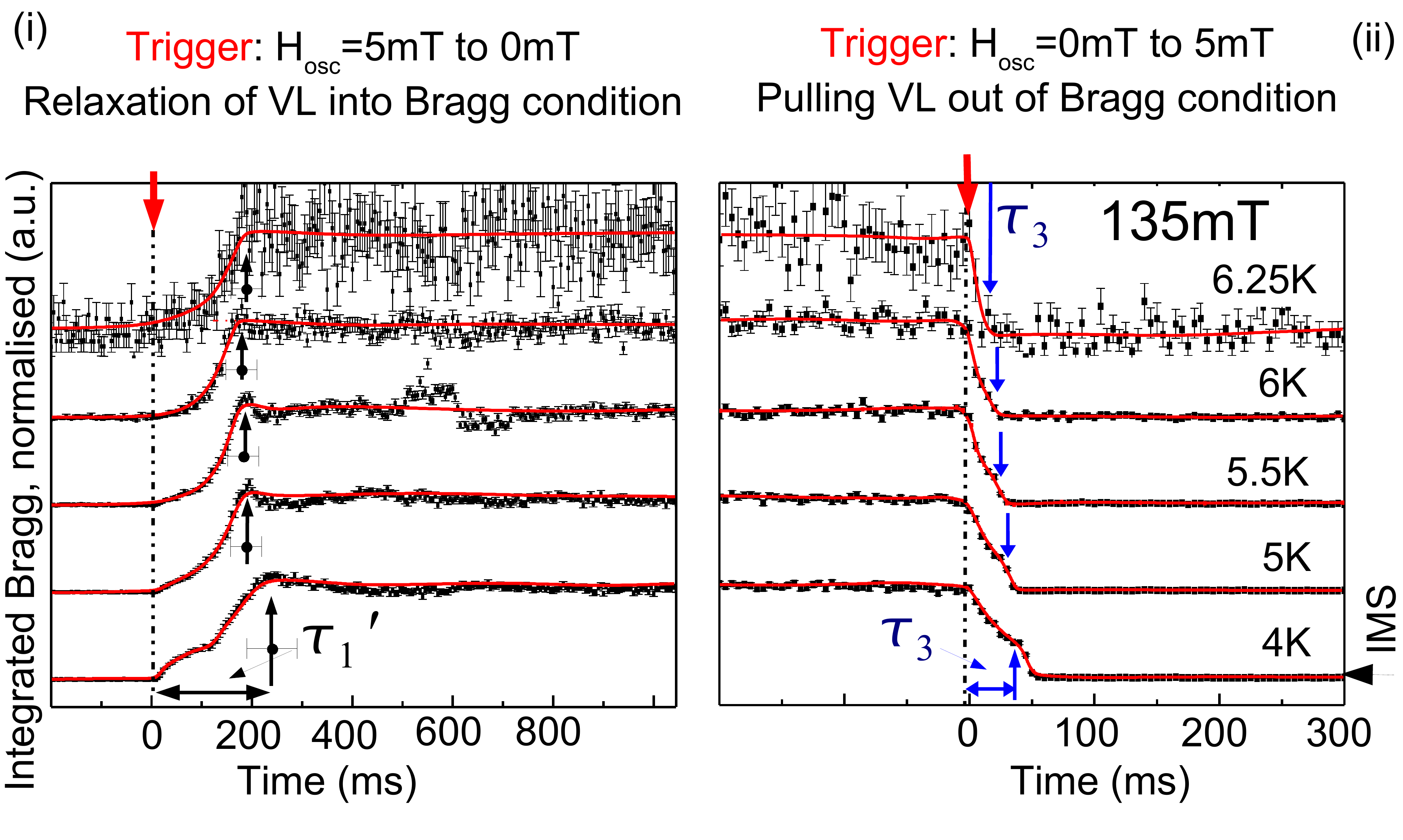}
\caption{Panel (i) shows {\it fixed angle scans} for $\mu_0H=$135\,mT analogous to Fig.\,\,\ref{tscans_1}.}
\label{tscans_3}
\end{center}
\end{figure}

\item The characteristic shape of the relaxation changes as a function of temperature and magnetic field: For the low temperature data points at each magnetic field, namely for $4\,{\rm K} \leq T \leq 6\,{\rm K}$ at $\mu_0H=75\,\rm mT$, for $4\,{\rm K} \leq T \leq 5\,{\rm K}$ at $\mu_0H=100\,\rm mT$ and for $T=4\,{\rm K}$ at $\mu_0H=135\,\rm mT$, the relaxation is characterized by a sharp kink, associated with a distinct increase of intensity immediately after the magnetic field is changed. This sharp increase is more pronounced at lowest temperatures. It is associated with the step-like broadening of the VL mosaic, as observed in the {\it time resolved mappings}, e.g., for $T=4\,{\rm K}$ and $\mu_0H=100\,\rm mT$ (Fig.\,\,\ref{bigpics_all}, panels (iii) and (iv)). The step-like increase is followed by a slow relaxation characteristic of the time-scale $\tau_1'$. Note, that the data points at low temperatures and low fields, where the kink is observed, are situated in the IMS. 

\item In contrast, the characteristic shape of the relaxation process exhibits a smooth increase with exponential shape for high temperatures, namely for $6.5\,{\rm K} \leq T \leq 7.25\,{\rm K}$ at a magnetic field of $\mu_0H=75\,\rm mT$, for $6\,{\rm K} \leq T \leq 6.75\,{\rm K}$ at $\mu_0H=100\,\rm mT$ and for $5\,{\rm K} \leq T \leq 6.25\,{\rm K}$ at $\mu_0H=135\,\rm mT$. This is attributed to the lack of the above-mentioned sharp increase of intensity directly after the magnetic field is changed.
\end{itemize}

We now concentrate on panels (ii) of Figs.\,\,\ref{tscans_1}, \ref{tscans_2} and \ref{tscans_3}. The decrease of intensity after the vertical line, labeled with {\it trigger}, is attributed to the VL being pulled {\it out} of the Bragg condition. The decreasing intensity is characterized by two different time-scales: $\tau_3$ characterizes the overall time after the change of magnetic field direction when the scattering intensity has decreased to $1/e$ of its initial value. $\tau_3$ is indicated with a blue marker in each scan. In contrast, $\tau_2$, indicated with a green marker, describes the characteristic delay-time between the change of magnetic field direction and the response of the VL. The temperature and magnetic field dependence of $\tau_2$ and $\tau_3$ is shown in Fig.\,\,\ref{diffusionconstants_3}, panel (ii):

\begin{itemize}
\item $\tau_3$ exhibits a linear decrease as a function of increasing temperature for all magnetic fields measured: For $\mu_0H=75\,\rm mT$, $\tau_3$ decreases from $\tau_3 =0.08\pm0.01\,\rm s$ at $T=4\,{\rm K}$ to $\tau_3=0.02\pm0.005\,\rm s$ at $T=7.25\,{\rm K}$. For $\mu_0H=100\,\rm mT$, $\tau_3$ decreases from $\tau_3=0.06\pm0.01\,\rm s$ at $T=4\,{\rm K}$ to $\tau_3=0.02\pm0.005\,\rm s$ at $T=6.75\,{\rm K}$ and for $\mu_0H=135\,\rm mT$, $\tau_3$ decreases from $\tau_3=0.045\pm0.01\,\rm s$ at $T=4\,{\rm K}$ to $\tau_3=0.02\pm0.005\,\rm s$ at $T=6.25\,{\rm K}$. No signature of a crossover from the IMS to the Shubnikov phase is observed in the temperature and magnetic field dependence of $\tau_3$.

\item The temperature dependence of $\tau_2$ is characterized by a linear decrease from $\tau_2=$0.047$\pm$0.01\,s at $T=4\,{\rm K}$ to $\tau_2=0$ at $T=7.25\,{\rm K}$ for a magnetic field of $\mu_0H=75\,\rm mT$. For a magnetic field of $\mu_0H=100\,\rm mT$, $\tau_2$ decreases from $\tau_2=$0.03$\pm$0.005\,s at $T=4\,{\rm K}$ to $\tau_2=0$ for temperatures above $T=6\,{\rm K}$. In contrast, for $\mu_0H=135\,\rm mT$, $\tau_2=0$ for all temperatures. Note, that $\tau_2 \neq 0$ in the IMS.
\end{itemize}

\section{Interpretation}
\label{interpretation}
For an interpretation of our data, we first review the VL diffusion for uniform tilt before we qualitatively describe the VL relaxation for a displacement of the magnetic field direction. Following this we calculate the temperature and magnetic field dependence of the VL tilt modulus $c_{44}$ using the VL diffusion model for uniform distortions. We finally consider the peculiarities of the relaxation process of the VL in the IMS.

\subsection{Vortex Lattice Elasticity and Diffusion for Uniform Distortions}
We have derived the characteristic relaxation rates $\Gamma_1$ and $\Gamma_2$ of VLs in section (\ref{VL_elasticity}) which are in the range of 10$^{-9}$ s$^{-1}$ for Nb. In our experimental setup, the magnetic field direction was oscillated with a frequency of 0.2\,Hz. Therefore, the associated relaxation process of the superconducting VL can be calculated in the ${\bf k}=0$ limit. For a change of the magnetic field direction, as used in our experimental setup, the relaxation process is essentially given by the VL tilt modulus $c_{44}({\bf k}=0)$. 

For uniform distortions, the VL tilt modulus $c_{44}= B H$ depends only on the applied magnetic field $H$ which is in local equilibrium with the equilibrium induction $B$ \cite{Brandt:95}. For the case that the spatially varying part of $B$ is smaller than the average value of $B$, the highly nonlinear equations of motion may be linearized. The response of the VL in a bulk sample (sample diameter $r>>\lambda'$) to a changed magnetic direction field may then be written as damped diffusion process \cite{Brandt:90}, using the diffusion equation derived by Kes \cite{Kes:89}. 

It is important to note that the VL initially responds to a change of applied magnetic field $H$ solely at the surface of the superconductor, as the magnetic field is screened by supercurrents from the bulk of the specimen (in particular by the Meissner domains in the IMS). Due to continuity conditions, the slope of the VL is slightly refracted at the surface of the sample. However, for the following description, the refraction of the magnetic field is neglected.

According to the diffusion model, the distortion of the VL propagates from the surface of the sample into the bulk due to the finite elastic constants of the VL. The resulting diffusion equation of the tilt distortion $u(x,t)$ of the VL is given by
\begin{equation}
\frac {{\rm d}u}{{\rm d}t}= D \cdot \frac {{\rm d^2}u}{{\rm d}x^2}
\end{equation}
with the diffusion constant $D$ given by the ratio of the tilt modulus $c_{44}$ and the viscosity $\eta$ \cite{Brandt_priv}:
\begin{equation}
D=\frac{c_{44}}{\eta}.
\label{c44}
\end{equation}
$\eta$ describes the viscous damping of the vortex motion by flux flow resistivity, assuming either vanishing pinning effects or an efficient depinning due to thermally assisted flux flow effects (cf. section (\ref{VL_elasticity}), eq. (\ref{damping})). This yields for the diffusion constant $D$:
\begin{equation}
D(T) =\frac{H \rho_n(T)}{B_{c2}(T)}
\label{D}
\end{equation}

The diffusion equation can be solved easily, if the sample is approximated by a flat plate with thickness $2r$, ignoring the pre-existing field $H_{ stat}$: We consider a conducting plate to which the transverse field $H_{\rm osc}$ is applied at $t=0$. The distribution of the field across the plate is then described through a square wave. As time progresses, the edges and then the middle of the sample relax to the outside applied field. This corresponds to the so-called Dirichlet condition of the diffusion equation which is generally solved by a sum of cosine waves with a half period of $2r$:
\begin{equation}
u(x,t)= \sum_{n=1}^{\infty} D_n(\cos{\frac{n \pi x}{2r}}) e^{\frac{-n^2\pi^2Dt}{4r^2}}
\end{equation}
where $n= 1,3,5...$ and
\begin{equation}
D_n= \frac{2}{2r} \int_0^{2r} f(x) \cos{\frac{n \pi x}{2r}} {\rm d}x
\end{equation}
with the initial condition $f(x,t=0)$.

The Fourier components thereby decay {\it independently} with a characteristic time $\propto 1/n^2$, so that after a short time, only the fundamental remains. This yields for the relaxation time for the mode with $n=1$
\begin{equation}
\tau = -\frac{4r^2}{D \pi^2}\,.
\end{equation}

For a sample of cylindrical shape and radius $r$ with the magnetic field applied perpendicular to the cylinder axis, the resulting diffusion constant for a {\it rotation} of the magnetic field with respect to the cylinder axis was calculated by Brandt \cite{Brandt:90}. The resulting relaxation time is
\begin{equation}
\tau_r \approx \frac{r^2}{(2.405)^2 D}
\label{diffusion_cylinder}
\end{equation}
where $x_0=2.405$ is the first node in the Bessel function $J_0(x)$.

\subsection{Vortex Lattice Relaxation Process}
We qualitatively describe the diffusion process of the VL. As the repetition cycle of $H_{\rm osc}=0.2$\,Hz is slow compared to the relaxation processes $\tau_1$, $\tau_1'$, $\tau_2$ and $\tau_3$ of the VL, a complete relaxation can be presumed for each measurement cycle of the stroboscopic measurement, i.e. the VL is in the same equilibrium state before each change of magnetic field direction.

We assume that the VL has relaxed for $\mu_0 H_{\rm osc}=5$\,mT. The VL thus does not satisfy the Bragg condition. No scattering intensity is observed. We then consider the next cycle of the stroboscopic measurement where $\mu_0 H_{\rm osc}=5$\,mT is decreased to $\mu_0 H_{\rm osc}=0$\,mT: The VL firstly interacts with the changed magnetic field direction at the surface of the sample. The perturbation then diffuses into the sample. The VL relaxates {\it into} the Bragg condition. The relaxation of the VL from the equilibrium position for $\mu_0H_{\rm osc}=5$\,mT (and thus $\epsilon =2^{\circ}$) to  $\mu_0 H_{\rm osc}=0$\,mT (and thus $\epsilon =0^{\circ}$) yields a large displacement $u\gg\lambda'$ of the vortices. This relaxation process is characteristic of a slow exponential relaxation with the timescales $\tau_1$ and $\tau_1'$ as defined in the previous section. The temperature and magnetic field dependence of $\tau_1$ and $\tau_1'$ was given in Fig\,\,\ref{diffusionconstants_3}, panel (i).

We now assume that the VL has relaxed for $\mu_0 H_{\rm osc}=0$\,mT, giving rise to maximum scattering intensity. If the direction of the magnetic field is shifted as $H_{\rm osc}$ is increased from $\mu_0 H_{\rm osc}=$0\,mT to 5\,mT, the VL again interacts with the changed magnetic field direction at the surface of the sample. The perturbation diffuses into the sample, the VL is pulled out of the Bragg condition, and the scattering intensity at the detector decreases. After $\tau_3$, the intensity has decreased to $1/e$ of its initial value. $\tau_3$ is thus a measure for the time-scale when the perturbation of the VL propagates across the complete sample \footnote{Strictly speaking, the perturbation propagates through the sample from both sides as the geometry is symmetric.}.

\subsection{Vortex Lattice Tilt Modulus $c_{44}$}
\begin{figure}
\begin{center}
\includegraphics[width=0.45\textwidth]{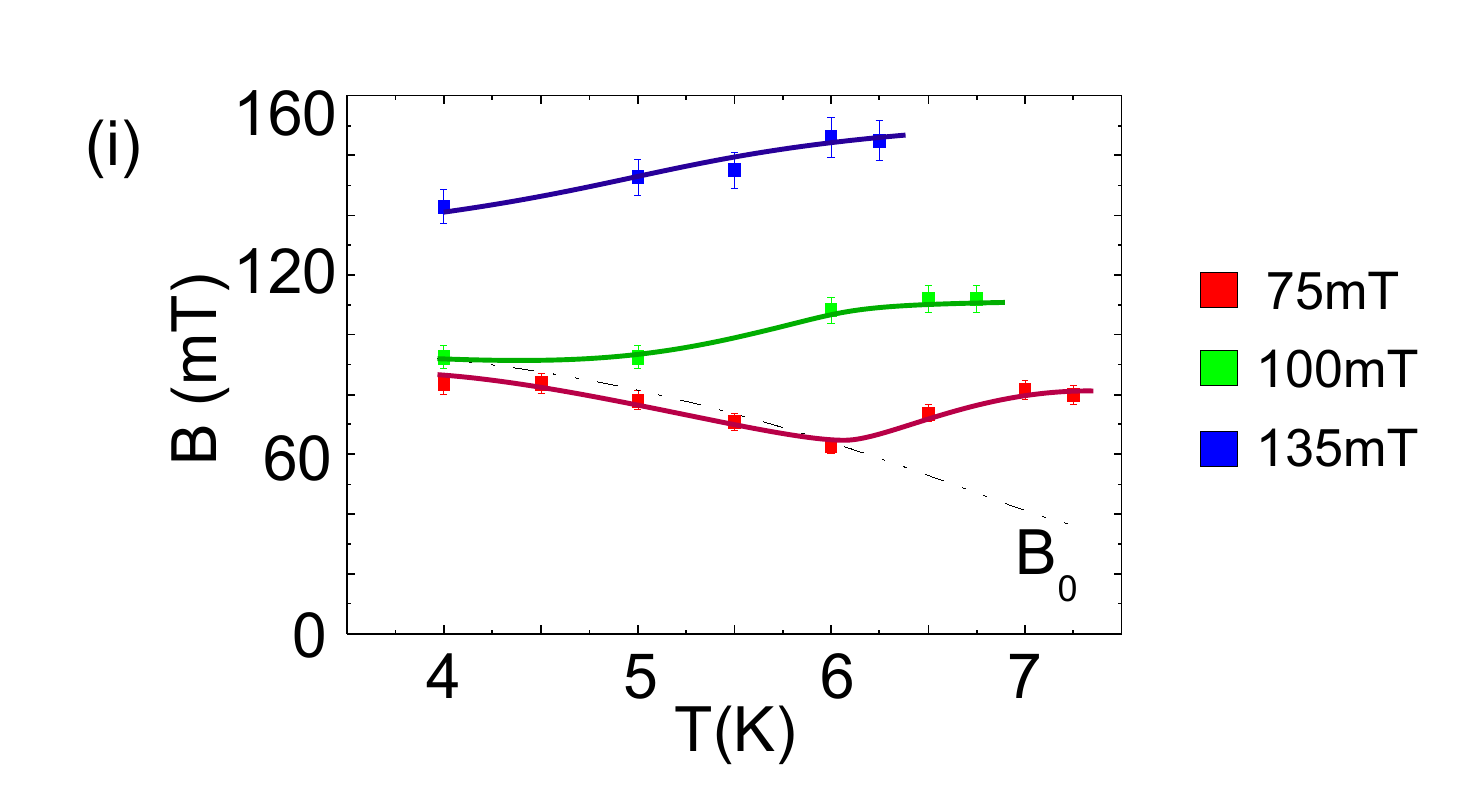}
\caption{Panel (i) depicts the temperature dependence of the equilibrium magnetization $B$ for $\mu_0 H=$ 75\,mT, 100\,mT and 135\,mT. The lines serve as guide to the eye.}
\label{diffusionconstants_4}
\end{center}
\end{figure}

According to the diffusion model for uniform distortions introduced by Kes \cite{Kes:89} and Brandt \cite{Brandt:95,Brandt:90}, the VL tilt modulus $c_{44}$ can be derived from $\tau_1$ and $\tau_3$ using the eqs.\,(\ref{c44}), (\ref{damping}), (\ref{D}) and (\ref{diffusion_cylinder}). The temperature dependence of $\rho_n(T)$ is given by eq.\,(\ref{rho}). The equilibrium induction $B$ is inferred from the modulus of the reciprocal lattice vector $|{\bf G}_{VL}|$ of the VL. Its temperature and magnetic field dependence is given in Fig.\,\,\ref{diffusionconstants_4}, panel (ii). 

\begin{figure}
\begin{center}
\includegraphics[width=0.45\textwidth]{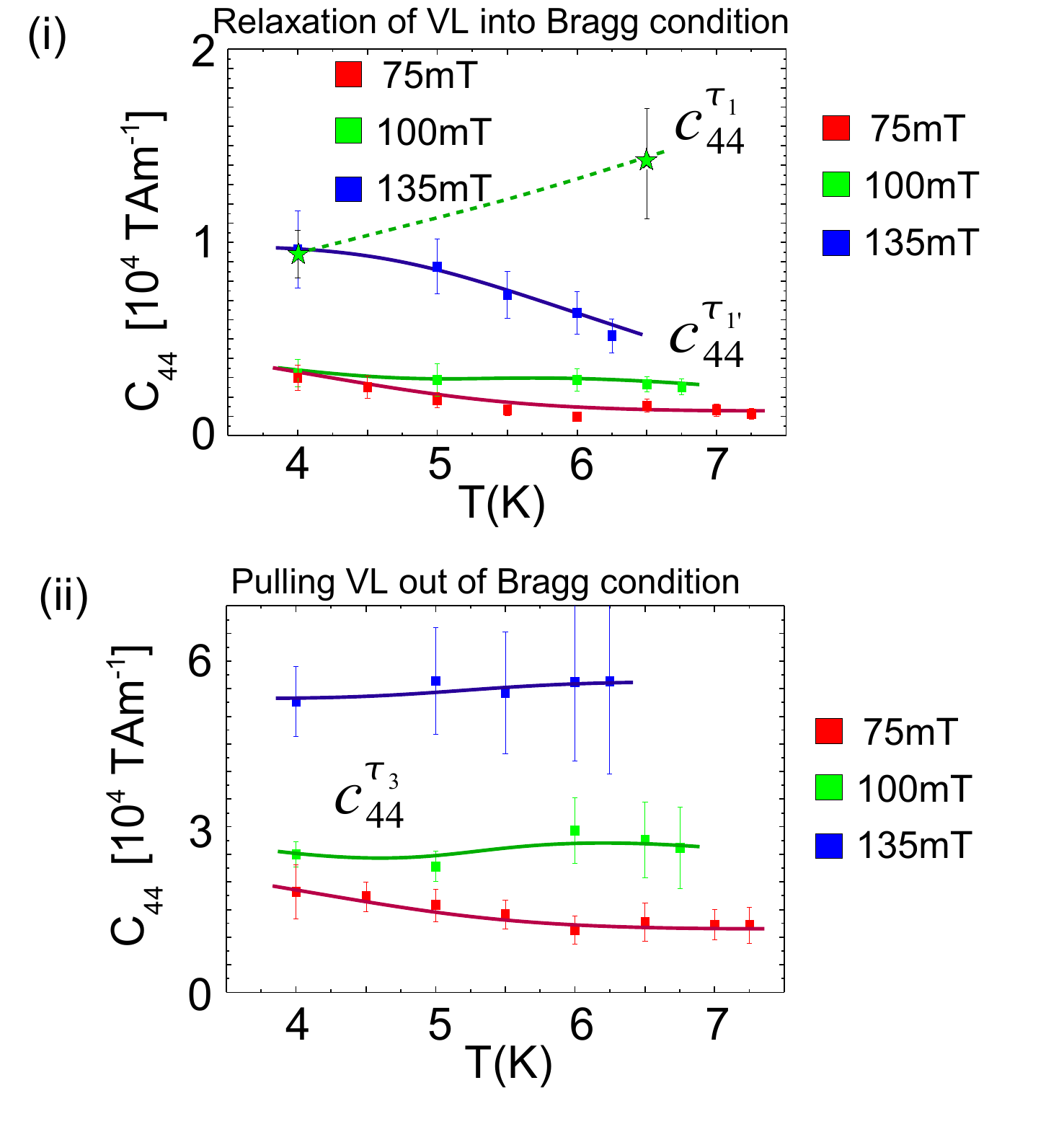}
\caption{Panels (i) and (ii) show the calculated temperature and magnetic field dependence of the VL tilt modulus $c_{44}^{\tau_1}(T,H)$, $c_{44}^{\tau_1'}(T,H)$ and $c_{44}^{\tau_3}(T,H)$, according to the model of Brandt \cite{Brandt:90}. The lines serve as guide to the eye.}
\label{diffusionconstants_2}
\end{center}
\end{figure}

The temperature and field dependence of $c_{44}^{\tau_1}(T,H)$, calculated from the measured values of $\tau_1$, as well as $c_{44}^{\tau_1'}(T,H)$ calculated from $\tau_1'$ and $c_{44}^{\tau_3}(T,H)$, calculated from $\tau_3$ is given in Fig.\,\,\ref{diffusionconstants_2}, respectively.

Only two data points could be obtained for $c_{44}^{\tau_1}(T,H)$ at $\mu_0H=100$\,mT for $T=4\,{\rm K}$ and $T=6.5\,{\rm K}$ from the {\it time resolved mappings}. $c_{44}^{\tau_1}$ shows increasing vortex stiffness with increasing temperature, increasing from $c_{44}^{\tau_1}\approx 1\cdot10^4$ TAm$^{-1}$ for $T=4\,{\rm K}$ to $c_{44}^{\tau_1}\approx 1.4\cdot10^4$ TAm$^{-1}$ for $T=6.5\,{\rm K}$.

The VL tilt modulus $c_{44}^{\tau_1'}(T,H)$ shows increasing VL stiffness with increasing magnetic field. Moreover, for magnetic fields $\mu_0H=$75\,mT and 135\,mT, $c_{44}^{\tau_1'}$ exhibits a weak decrease by a factor of two for increasing temperature from $T=4\,{\rm K}$ to $T_c$. In contrast, for $\mu_0H=$100\,mT $c_{44}^{\tau_1'}=0.3\cdot10^4$ TAm$^{-1}$ shows no temperature dependence. For a temperature of $T=4\,{\rm K}$ and a magnetic field of $\mu_0H=$135\,mT $c_{44}^{\tau_1'}$ yields $c_{44}^{\tau_1'}\approx 1\cdot10^4$ TAm$^{-1}$.  For a temperature of $T=4\,{\rm K}$ and magnetic fields of $\mu_0H=$75\,mT and 100\,mT the VL tilt modulus $c_{44}^{\tau_1'}\approx 0.3\cdot10^4$ TAm$^{-1}$. 

$c_{44}^{\tau_3}(T,H)$ also shows increasing VL stiffness with increasing magnetic field. However, no significant temperature dependence is observed. Due to $\tau_1' \gg \tau_3$,  $c_{44}^{\tau_3}$ yields $c_{44}^{\tau_3}\approx 5.5\cdot10^4$ TAm$^{-1}$ for $\mu_0H=$135\,mT, $c_{44}^{\tau_3}\approx 2.5\cdot10^4$ TAm$^{-1}$ for $\mu_0H=$100\,mT, and $c_{44}^{\tau_3}\approx 2\cdot10^4$ TAm$^{-1}$ for $\mu_0H=$75\,mT.

\begin{figure}
\begin{center}
\includegraphics[width=0.45\textwidth]{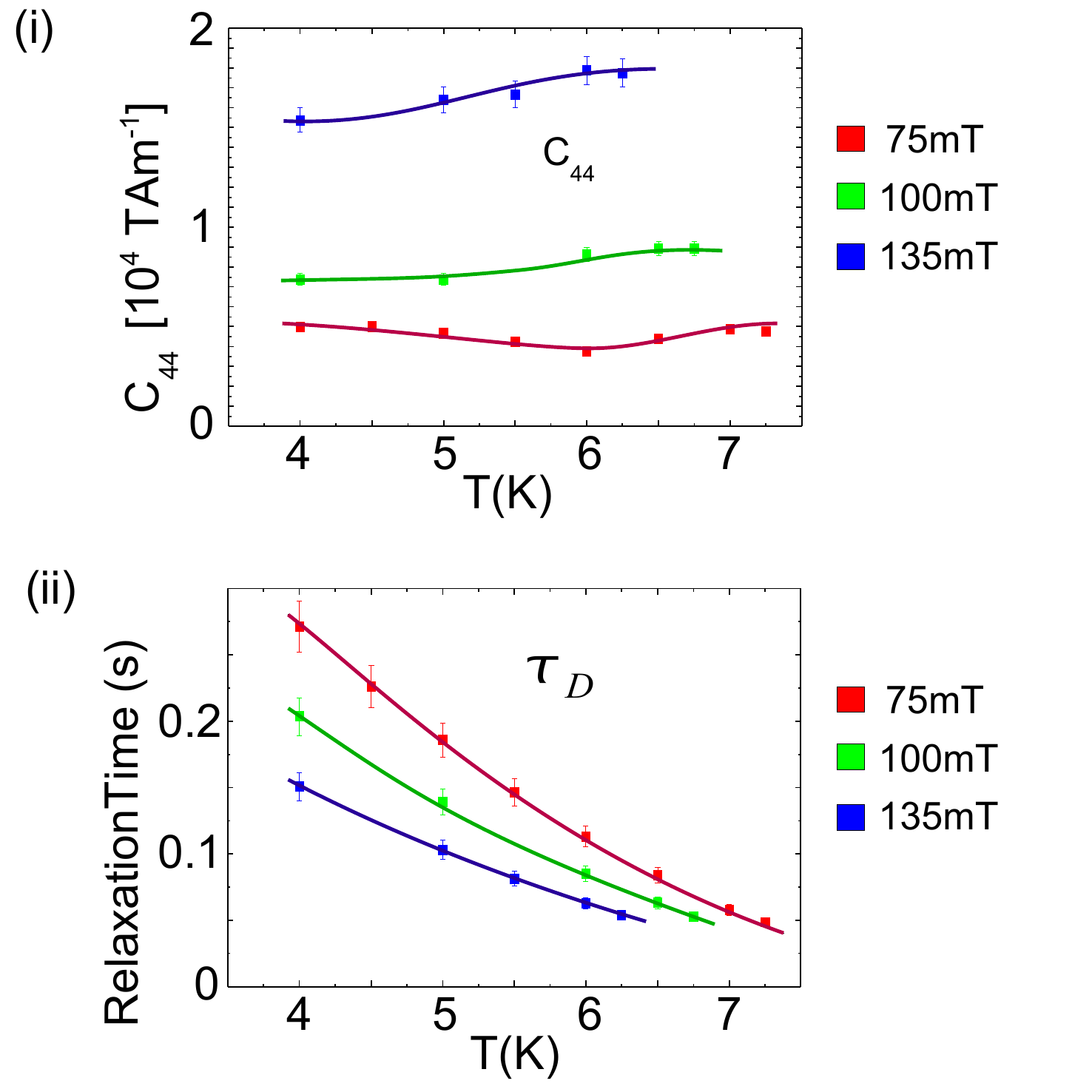}
\caption{The VL tilt modulus $c_{44}$ calculated from literature values is plotted in panel (i), whereas panel (ii) yields the calculated timescale for the diffusion of a VL distortion in a cylindric Nb sample, according to the model of Brandt \cite{Brandt:90}. The lines serve as guide to the eye.}
\label{diffusionconstants_1}
\end{center}
\end{figure}

The expected temperature and magnetic field dependence of $c_{44}$, calculated from literature values for Nb is given in Fig.\,\,\ref{diffusionconstants_1}, panel (i). The calculated diffusion time $\tau_D$ is given in panel (ii). The effects of thermal depinning have been neglected.  For a magnetic field of $\mu_0H=$135\,mT $c_{44}$ assumes a value of 1.7$\cdot10^4$\,TAm$^{-1}$, for $\mu_0H=$100\,mT $c_{44}=$0.8$\cdot10^4$\,TAm$^{-1}$ and finally for $\mu_0H=$75\,mT $c_{44}=0.6\cdot10^4$\,TAm$^{-1}$. The temperature dependence of $c_{44}$ is weak for all fields and reflects the temperature dependence of the equilibrium induction $B$.

Qualitatively the measured values of $c_{44}^{\tau_1}(T,H)$, $c_{44}^{\tau_1'}(T,H)$ and $c_{44}^{\tau_3}(T,H)$ compare well with the theoretical value of $c_{44}(T,H)$: Firstly, the magnetic field behaviour is consistent with an increased VL stiffness $c_{44}$ for increasing field, according to eq. (\ref{elasticity}). The most accurate agreement is obtained for $c_{44}^{\tau_1}(T,H)$. The origin of the deviation is most likely due to uncertainties of the extrapolation of the normal conducting resistivity $\rho_n$ to low temperatures, associated with the extrapolation of the RRR. The values of $c_{44}^{\tau_1'}(T,H)$ are lower by a factor two from the expected values. In contrast, the values for $c_{44}^{\tau_3}(T,H)$ exceed the calculated values by a factor of 3.5. This deviation is presumably resulted by the definition used for $\tau_1'$ and $\tau_3$.

We have introduced in the previous paragraphs that the Fourier modes of the VL relaxation decay independently with a time constants $\propto 1/n^2$. Note, that the values of $\tau_1$ exceed $\tau_3$ by a factor of approximately two to three. Whereas the fundamental relaxation mode $\tau_1$ may be identified with the Fourier mode for $n=1$ \footnote{After a certain time, only the fundamental mode survives.}, the values obtained for $\tau_3$ suggest that the corresponding relaxation process is a mixture of several different mechanisms. Moreover, the involvement of the shear modulus $c_{66}$ in the relaxation process is still unclear. The shear modulus is responsible for the perfection of the {\it local structure} of the VL. However, no azimuthal smearing of the scattering pattern was observed {\it during} the relaxation process of the VL, indicating a disordering akin to a melting transition of the VL. This leads to the following picture of the VL relaxation, already accounting for most the observed features:

(i) In general, a faster propagation of perturbations in the VL is observed for increasing temperatures and increasing magnetic fields. This general behaviour is explained by the decreased damping $\eta$ for increasing temperature, according to an increase of $\rho_n(T)$ with $T$ as given in eq. (\ref{damping}) and increasing VL stiffness $c_{44}$ with increasing field according to $c_{44}= B H$. 

(ii) The {\it time resolved mappings} show that at high temperatures and high magnetic fields in the Shubnikov phase, the VL responds to the changed magnetic field direction as a single, stiff lattice with its motion characterized by the diffusion constant $D$. A single, exponential relaxation is thus observed. 

(iii) In contrast, due to the low stiffness and strong damping for low temperatures and fields in the IMS, the macroscopic relaxation is typical of a slow exponential relaxation with large mosaic spread which is characterized by an additional fast process on a short timescale. This is attributed to the decomposition of the VL into Shubnikov domains and Meissner phase domains in the IMS. The origin of the additional fast process will be discussed below.

(iv) The macroscopic relaxation of the VL between the two equilibrium positions is associated with a large vortex displacement. It is strongly dependent on the VL topology, as the crossover from the IMS to the Shubnikov phase is reflected in the mere temperature dependence of $\tau_1'$.

(v) $\tau_3$ is a measure for the time of a VL distortion propagating through the sample. No signature of the transition from IMS to the Shubnikov phase is observed in the temperature and magnetic field dependence of $\tau_3$ which is thus insensitive to the VL topology.

\subsection{Vortex Lattice Relaxation in the Intermediate Mixed State}

\begin{figure}
\begin{center}
\includegraphics[width=0.35\textwidth]{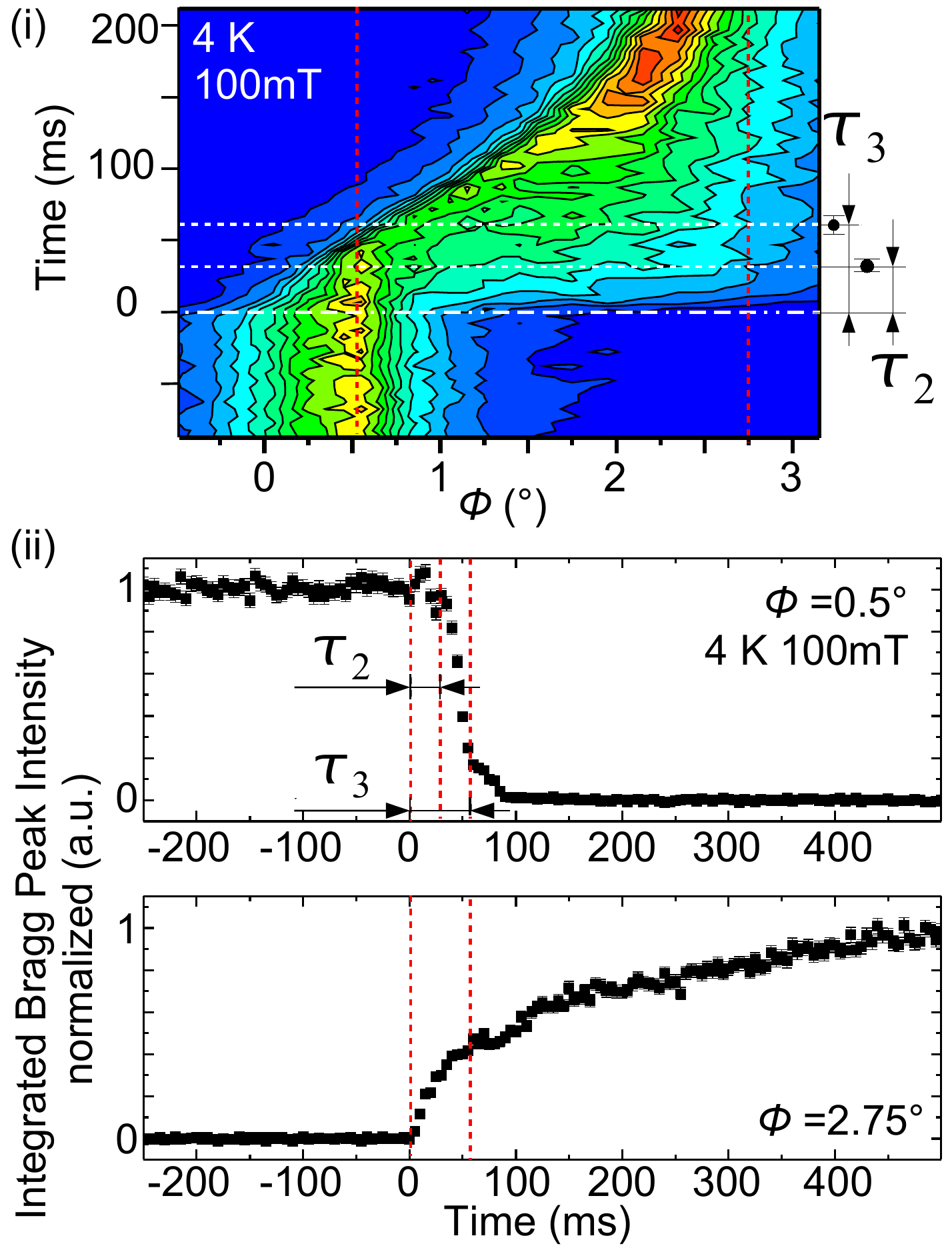}
\caption{Panel (i): {\it time resolved mapping} for $T=$4\,K and $\mu_0H=$100\,mT. The relaxation process of the VL is shown for a short time-scale after the magnetic field direction was switched. Panel (ii) shows the associated {\it fixed angle scans} for $\phi=0.5^{\circ}$ and $\phi=0.2.75^{\circ}$ as indicated with the vertical broken red lines in panel (i).}
\label{switching}
\end{center}
\end{figure}

Certain characteristic features show up exclusively for data points in the IMS. Typical data of the relaxation process for $T=4\,{\rm K}$ and $\mu_0H=$100\,mT is shown in Fig.\,\,\ref{switching}. The time range displayed corresponds to a short time-scale after the magnetic field direction was changed. In the {\it time resolved mapping} (panel (i)), a sharp increase of mosaicity shows up which is associated with a sharp increase of intensity observed in the {\it fixed angle scans} for the equilibrium position at $\phi=2.75^{\circ}$ (panel (ii)). We note, that the characteristic time-scale of this feature is well below $\tau_3$ for all temperatures and fields in the IMS. 

We have introduced $\tau_3$ as a measure for the time, necessary for a perturbation of the VL to cross the sample. This locates the related relaxation process responsible for the sharp increase of intensity and mosaicity at the surface of the sample. This effect is hence attributed to branching of the Shubnikov domains at the surface of the sample in the IMS. The branching of Shubnikov domains leads to a fine VL structure at the surface, consisting of connected Shubnikov domains with open topology enclosing Meissner islands. In particular, no rigid VL is observed. Branching of the VL is responsible for the large intrinsic mosaicity as well.

As remaining feature, we now discuss the characteristic delay $\tau_2$ when the VL is {\it pulled out} of the Bragg condition. A delay $\tau_2=$30\,ms is visible in both the {\it time resolved mapping} (panel (i)) and the corresponding {\it fixed angle scan} for $\phi=0.5^{\circ}$ (panel (ii) of Fig.\,\,\ref{switching}). Note, that $\tau_2\neq 0$ solely for data points in the IMS. A sharp increase of intensity is observed simultaneously at the new equilibrium position ($\phi=2.75^{\circ}$) for these measurement points. This sharp increase was attributed to a mechanism close to the surface of the sample. The IMS is characterized a bending of vortices at the surface. The scattered intensity {\it exactly} at the position ${\bf q}={\bf G_{VL}}={\bf k}_i -{\bf k}_f$ is caused by the VL which points to the initial equilibrium position, thus which is {\it not} bent due to branching. $\tau_2$ thus describes the delayed reaction of the unbranched VL buried beneath the surface of the sample.

\section{Conclusions and Outlook}
\label{conclusion}

In our study we measure the dynamic properties of bulk VLs with time resolved stroboscopic SANS combined with a time varying magnetic field setup. We show that the qualitative magnetic field and temperature dependent behaviour of the VL in the Shubnikov phase for uniform distortions can be described in reasonable agreement with a diffusion model by Brandt \cite{Brandt:90, Brandt:95} and Kes \cite{Kes:89}. The values obtained for the VL tilt modulus $c_{44}$, diffusivity $D$ and damping $\eta$ can be reproduced theoretically. We further argue that the topology of the VL is reflected sensitively in the associated diffusion process. This is readily seen for measurement points in the IMS where a second, very fast relaxation process maybe attributed to branching of the VL domains at the surface of the sample. Such a branching was observed by previous high resolution decoration techniques on bulk Nb single crystal samples \cite{Golubok:82, Essmann:66, Brandt:10}. The effect of surface treatment and roughness on the details of the Landau branching remains as a subject of future studies.

Our study represents a experimental technique for how to access directly VL melting and the formation of vortex glass states in unconventional superconductors, notably the cuprates, heavy-fermion systems, borocarbide or iron arsenide systems. The possibility to precisely determine the pinning properties of vortices in future experiments on samples of varying purity is of great relevance for the research on technical applications of superconducting devices, as the pinning properties are intimately related to the maximal critical current density of superconductors. 

Furthermore, the technique developed for our study is of general relevance for materials exhibiting complex forms of magnetic order, i.e. long range helical order as observed in materials without inversion symmetry \cite{Pfleiderer:09, Muenzer:09}, magnetic Skyrmion lattices, as observed recently in the helimagnet MnSi \cite{Muehlbauer:09b} or colloidal magnetic suspensions, liquid crystals and also Bose condensates and glasses of magnetic triplet excitations in quantum magnets \cite{Hong:2010}.

To further increase the time resolution, the TISANE technique can be used instead of stroboscopic small angle neutron scattering \cite{Kipping:08}. TISANE benefits from a neutron chopper which is placed upstream of the sample position at the distance $L_1$ to the sample. By carefully adjusting the distances $L_2$ and $L_1$, the chopper and control parameter duty cycle frequencies and phase, a coherent summation of the scattered neutrons at the detector position in {\it space and time} without time smearing can be achieved. The accessible time resolution of TISANE is mainly determined by the time resolution of the detector and the opening time of the chopper system. Timescales from $\mu$s to several ms are possible, closing the gap to the inelastic technique NRSE. For a detailed description, we refer to \cite{Wiedenmann:06, Kipping:08}.

We wish to thank R. Hackl, M. Laver, M. Janoschek, T. Adams, U. Essmann, R.Georgii, W. Zwerger, M. Garst and A. Zheludev for support and stimulating discussions. Technical support and help from R. Schwikowski, B. Russ, D. Wallacher and P. Granz is greatfully acknowledged.

\end{document}